\documentclass[sigconf,authorversion,preprint]{acmart}

\AtBeginDocument{%
	}

\copyrightyear{2024} 
\acmYear{2024} 
\setcopyright{acmlicensed}\acmConference[ETRA '24]{2024 Symposium on Eye Tracking Research and Applications}{June 4--7, 2024}{Glasgow, United Kingdom}
\acmBooktitle{2024 Symposium on Eye Tracking Research and Applications (ETRA '24), June 4--7, 2024, Glasgow, United Kingdom}
\acmDOI{10.1145/3649902.3653521}
\acmISBN{979-8-4007-0607-3/24/06}

\usepackage{varwidth}

\newcommand{\wsg}[1]{\raisebox{-3.0pt}[0pt][0pt]{\includegraphics[height=1.2em]{#1}}}
\newcommand{\wsgLabelOne}[1]{#1}
\newcommand{\wsgLabelTwo}[1]{#1}
\newcommand{\wsgLabelThree}[1]{#1}

\usepackage{subcaption}
\usepackage{xspace} 
\newcommand{\base}{\wsg{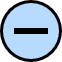}\xspace}
\newcommand{\baseTwo}{\wsg{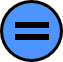}\xspace}
\newcommand{\highlight}{\wsg{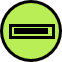}\xspace}
\newcommand{\icon}{\wsg{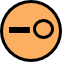}\xspace}
\newcommand{\wsgCond}{\wsg{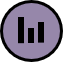}\xspace}

\newcommand{\added}[1]{#1}
\newcommand{\removed}[1]{}

\begin{document}
	
	\title{Eye Tracking on Text Reading with Visual Enhancements}
	
	\author{Franziska Huth}
	\email{franziska.huth@vis.uni-stuttgart.de}
	\affiliation{%
		\institution{University of Stuttgart}
		\country{Germany}
	}
	\orcid{0000-0003-1393-5641}
	
	\author{Maurice Koch}
	\email{maurice.koch@visus.uni-stuttgart.de}
	\orcid{0000-0003-0469-8971}
	\affiliation{%
		\institution{University of Stuttgart}
		\country{Germany}
	}
	
	\author{Miriam Awad}
	\email{miriam.awad@hotmail.de}
	\affiliation{%
		\institution{University of Stuttgart}
		\country{Germany}
	}
	\orcid{0009-0000-0315-1697}
	
	\author{Daniel Weiskopf}
	\email{weiskopf@visus.uni-stuttgart.de}
	\orcid{0000-0003-1174-1026}
	\affiliation{%
		\institution{University of Stuttgart}
		\country{Germany}
	}
	
	\author{Kuno Kurzhals}
	\email{kuno.kurzhals@visus.uni-stuttgart.de}
	\orcid{0000-0003-4919-4582}
	\affiliation{%
		\institution{University of Stuttgart}
		\country{Germany}
	}

	\renewcommand{\shortauthors}{Huth, et al.}
	
	\begin{abstract}
		The interplay between text and visualization is gaining importance for media where traditional text is enriched by visual elements to improve readability and emphasize facts.
		In two controlled eye-tracking experiments ($N=12$), we approach answers to the question:
		How do visualization techniques influence reading behavior?
		We compare plain text to that marked with highlights, icons, and word-sized data visualizations.
		We assess quantitative metrics~(eye movement, completion time, error rate) and subjective feedback~(personal preference and ratings).
		The results indicate that visualization techniques, especially in the first experiment, show promising trends for improved reading behavior.
		The results also show the need for further research to make reading more effective and inform suggestions for future studies.
		
	\end{abstract}

	\ccsdesc[500]{Human-centered computing~Empirical studies in visualization}

	\keywords{Eye tracking, reading, visualization, visual highlighting, word-sized graphics}
	
	
	\maketitle
	
	\section{Introduction}
\label{sec:introduction}
The investigation of reading behavior has a long tradition in psychology research~\cite{huey1908}.
For eye tracking research, the seminal work of \citet{rayner1977} helped identify common eye movements during reading and information processing.
Since then, many experiments have been conducted to identify how people read and interpret different visual stimuli~\cite{rayner1998eye}.
One important category of such stimuli is data visualizations and infographics, often containing text and graphics to communicate information~\cite{goldberg2011, netzel2017}.

A simple way of enhancing text with visual elements relies on highlighting specific words and passages, for instance, to emphasize important facts during studies and to guide attention to results of keyword searches in text documents. 
Alternatively, symbolic representations of keywords can be embedded into text.
Early examples can be found in the book \textit{The Elements of Euclid} by \citet{byrne1847first}, where geometric shapes were integrated into the text of the book \textit{``... for the greater ease of learners.''}
Modern publications also use this technique, for instance, to differentiate between experimental conditions~\cite{9908291}.
Furthermore, visualization research investigated how to encode data in word-sized graphics to communicate information, e.g., temporal changes with sparklines~\cite{tufte_beautiful_2006}.
Those word-sized graphics have been successfully used, for example, to augment texts that explain eye-tracking data~\cite{beck2016expert,beck_word-sized_2017-1}, as part of scientific texts~\cite{latif_visually_2018}, or for better awareness of discussion topics in social media feeds~\cite{huth_word-sized_2021,huth2021online}.
\citet{goffin_exploring_2015,goffin_exploratory_2017} discussed design considerations of word-sized graphics in text and performed a study that finds positive effects on information gain.

A considerable body of work exists on text highlighting and its effects on comprehension and memory~\cite{fowler1974effectiveness, lorch1989text, ben2018contrib}, but
only a few eye-tracking studies investigated the effect of text highlighting on reading behavior.  
\citet{chi2007visual} found validation for the \emph{von Restorff isolation effect}~\cite{restorff1933wirkung}, which states that highlighted areas draw the attention of the reader, whether they are important or not.
Their study comprised three conditions; No highlights, keyword highlights, and ScentHighlights~\cite{chi2005scent}, which is an automatic approach for highlighting relevant keywords in texts. 
They found that approximately half of the fixations were in highlighted areas, suggesting a strong effect on reading behavior.
Similarly, \citet{yeari2017effect} studied if the level of information centrality, which refers to the importance of text segments, impacts online processing and offline memory.
This specific study found that attention is driven by both highlights and the level of centrality, meaning that readers directed their attention to important words even if they were not highlighted.
Interestingly, highlighting had no significant effect on the ability to recall information acquired by reading the text.
These studies only looked at text highlighting, while our work also investigates reading behavior on text with icons and word-sized graphics.
So far, there is no consensus on whether those visual enhancements aid people in recalling information from text documents ~\cite{ruchikachorn2018eye} or if they distract their reading flow.
In this work, we aim to fill this research gap by addressing the following question:
\vspace{1ex}
\begin{center}
	\mbox{\begin{varwidth}{\dimexpr\linewidth-30\fboxsep-2\fboxrule\relax}
			\textit{How do visual enhancements (i.e., highlighting, icons) and word-sized graphics influence reading behavior with respect to gaze patterns, reading speed, and information assimilation?}
	\end{varwidth}}
\end{center}
\vspace{1ex}
We hypothesize that there are differences in the reading behavior of visually enhanced text versus plain text, and that task performance increases, i.e., people can grasp more information quicker.
Our contributions in this work are two experiments to explore differences in viewing behavior when reading text with different visual enhancements.
We investigate deviations from the baseline (plain text) for a memory task of facts in a short description text.
Our results are also available via OSF~\cite{osf} and indicate differences in viewing behavior, especially for texts with highlighted words.

	\section{Eye Tracking of Visually Enhanced Text}
To find out how visual enhancement of text influences reading behavior, we conducted a small-scale study comprising two experiments investigating the influence of (1) highlighting and symbols and (2) word-sized graphics.

\subsection{Participants and Experimental Setup}
The experiments were conducted in a controlled lab environment isolated from outside distractions.
For remote eye tracking, we used a Tobii Pro Spectrum with a resolution of 1920\,$\times$\,1080 pixels.
Gaze was recorded at 1200\,Hz and processed into fixations by the Tobii I-VT filter with default settings.
We used a 9-point calibration and displayed the stimuli horizontally and vertically centered.

We invited $12$ participants who volunteered to perform both experiments in consecutive order. 
Participation was completely voluntary without further compensation.
$3$ of the participants were \wsgLabelThree{between 18 and 24}, $7$ \wsgLabelTwo{between 25 and 35}, and $2$ participants were \wsgLabelOne{older than 35}. 
$4$ participants identify as \wsgLabelTwo{female}, and $8$ as \wsgLabelOne{male}. 
All but one participant had good or very good English language skills.
Most participants frequently used icons or emojis and had high or very high experience with bar charts, line charts, and pie charts.
After giving their informed consent and answering some demographic questions, participants were introduced to each experiment.
They conducted one test task with a short text for each condition, before starting the experiment recording.
The two experiments took about $60$ minutes together per participant. 

We measured task performance as accuracy in answering the questions for each stimulus, time to read the texts and time to answer the questions.
In terms of completion times, we did not limit the time for participants to read the text, nor when asked the respective question to remember the facts of a text.
We also measured gaze data with fixations and saccades.
In both experiments, participants rated task difficulty, confidence in their answer, and their stress level when solving the task after each stimulus-question pair.
At the end of each experiment, we asked about participant's reading strategies and preferences regarding the visualization techniques.
All stimuli and questions can be found in this paper's OSF repository~\cite{osf}.

\subsection{Experiment 1: Highlighting and Symbols in Text}
The first experiment compared three conditions: (1) plain text (as baseline~\base), (2) highlighting~\highlight, and (3) symbols~\icon.
In a within-subjects design, we tested text augmented with each visualization type against the baseline without visual enhancements.
Highlighting specific terms in a text can steer readers' attention to important aspects and may help them grasp information quickly.
Similarly, symbols that augment important terms catch the attention to these terms and may ease text understanding.
Our goal was to measure the strength of those effects.

\begin{figure}
	\centering
	\includegraphics[width=\linewidth,trim={0 0 0 1mm}]{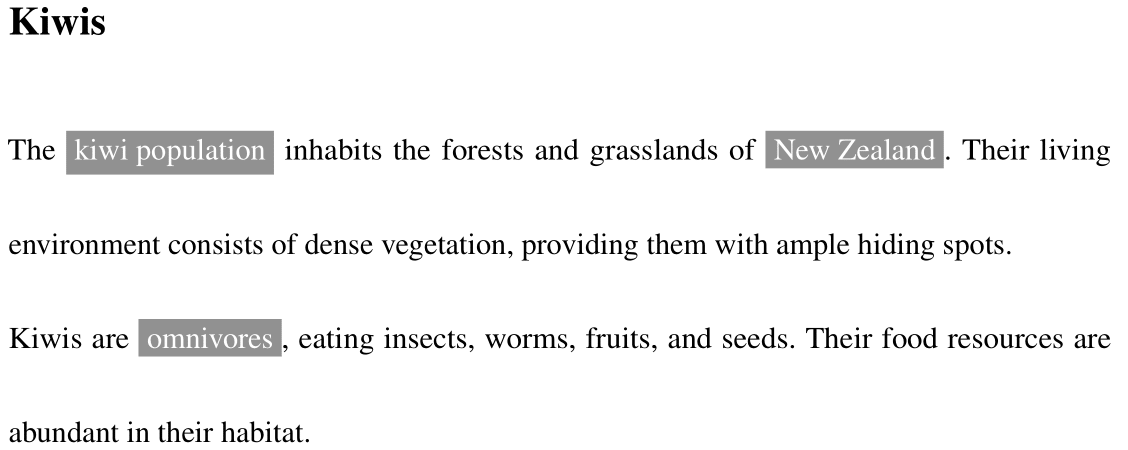}
	\hfill
	\hrule
	\vspace{1em}
	\includegraphics[width=\linewidth]{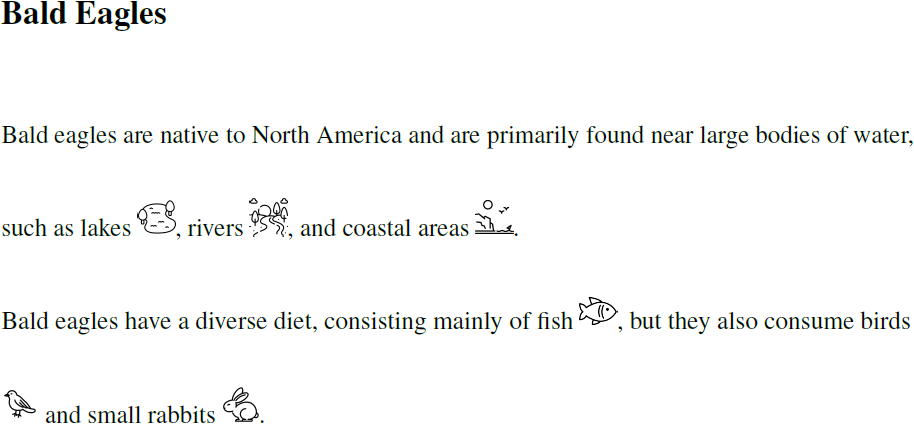}
	\caption{Excerpts from stimuli for the experiment conditions \textit{highlight}~\highlight (top) and \textit{icons}~\icon (bottom).}
	\label{fig:stimuli_exp1}
	\Description{This figure shows two images of text about animals.
		The image on top is a text about kiwis, with some of the factual terms in the text highlighted with a grey background, for example, ``New Zealand'' as a habitat region.
		The bottom text is about bald eagles.
		After some of the factual terms, there is an icon that depicts those terms.}
\end{figure}

\paragraph{Stimuli and Task}
We created short texts with 
\citet{gpt} about different animals in the style of a lexicon article.
The texts contained information about habitats, food sources, etc.
Some examples are depicted in Figure~\ref{fig:stimuli_exp1}.

In pilot studies within our lab, we iteratively refined the text length to avoid fatigue throughout the experiment, but still have enough information to challenge short-term memory.
We edited the texts to the same formatting and length of $95$ to $115$ words, so that each stimulus contained a similar number of facts about the respective animal.
We evenly emphasized about $50$ percent of these facts with the visual enhancements so participants could not rely on learning which type of fact was asked.
Because the facts in the texts were AI-generated, we warned the participants that they may not be entirely correct.
This feature also helped to balance against prior knowledge that participants might have had about the animals.
The participants were given five texts per condition, with the order of the conditions based on a \textit{balanced Latin square}.
Before each text appeared on the screen, we briefly showed an ``X'' at the top left to guide the gaze to the beginning position of the text.
When participants indicated they were done reading a text, we switched the view to the respective question about the text, in which we asked about one of the facts.
There were multiple correct answer options, for example, correct answers to ``What makes bald eagles skilled hunters?'' could be ``
good eyesight'' or ``diving abilities''.


\subsection{Experiment 2: Word-sized Graphics}
The second experiment considers how participants read text containing word-sized graphics.
Such visualizations depict data and are more complex than symbols.
Consequently, people have to investigate such visual elements in text longer and more intensely than symbols and highlighting, where preattentive visual features are often sufficient to convey their purpose.
Analogous to Experiment~1, Experiment~2 was performed in a within-subjects design with the same participants.

\begin{figure}
	\centering
	\includegraphics[width=\linewidth]{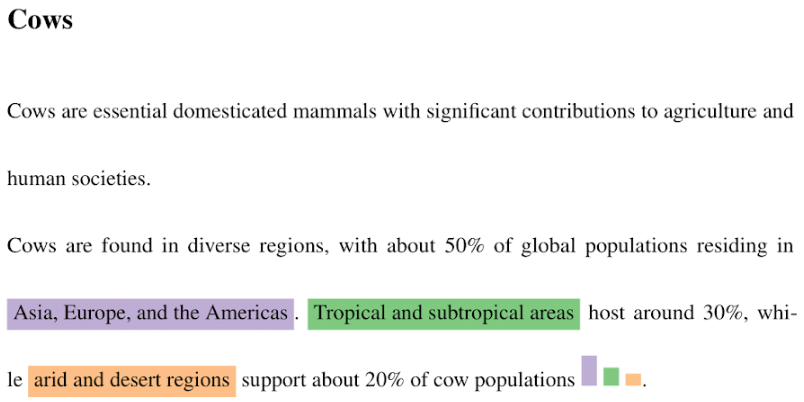}
	\hfill
	\hrule
	\vspace{1em}
	\includegraphics[width=\linewidth,trim={0 0 0 2.5mm}]{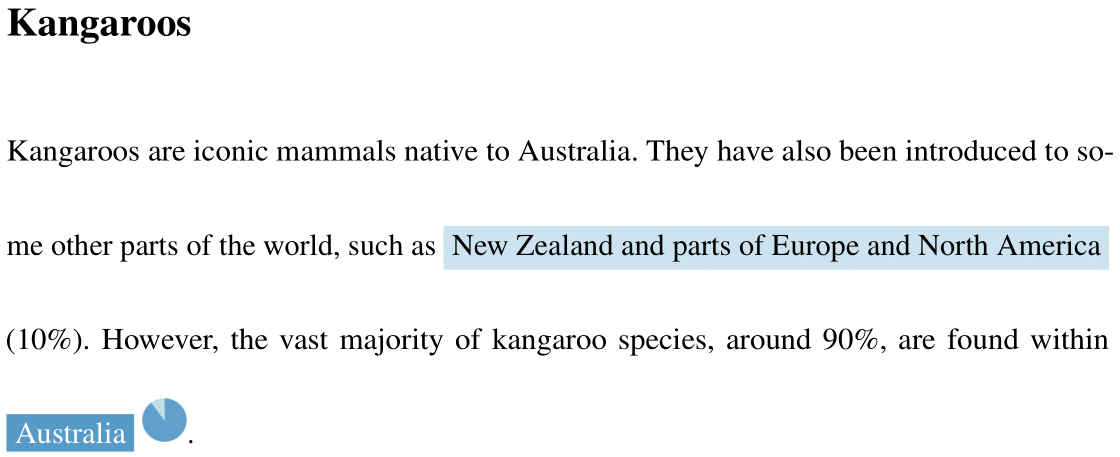}
	\caption{Excerpts from the stimuli for the experiment condition \textit{word-sized graphics}~\wsgCond.}
	\label{fig:stimuli_exp2}
	\Description{This figure shows two images of text about animals.
		The image on top is a text about cows, with some of the factual terms in the text highlighted with a colored background, for example, ``Asia, Europe, and the Americas'' as a habitat region.
		At the end of the sentence that has the colored terms, there is a bar chart that depicts the percentages of how many cows live in that region.
		The bottom text is about kangaroos.
		Again, some of the factual terms in the text are highlighted with a colored background.
		In this text, there is a pie chart that depicts the percentages.}
\end{figure}

\paragraph{Stimuli and Task}
We only included one condition in addition to the second baseline~\baseTwo of plain text.
The text was created similarly to Experiment~1, including statistical information enhanced by word-sized graphics~\wsgCond~(Figure~\ref{fig:stimuli_exp2}).
We added simple diagrams, such as bar charts, line charts, and pie charts, and highlighted the terms that correspond to the elements of the chart in the respective colors.
Again, about $50$ percent of the facts in each text were emphasized.
We chose basic visualization types because of their widespread familiarity and ease of understanding.
We gave each participant five texts, again in balanced order of conditions.
Similar to Experiment~1, we asked participants to read the text, and when they were done, we asked about one of the facts in the text.
Here, we asked about percentages, for example: ``How much of the total cow population inhabit tropical and subtropical
areas?'', to which the correct answer was ``$30\%$.''
We excluded one of the stimuli in the evaluation, because only after the experiment we noticed that there was no highlighted text that corresponded to the visualization that we asked about.

\begin{figure}
	\centering
	\includegraphics[width=\linewidth]{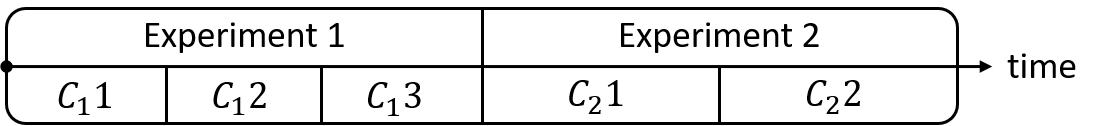}
	\caption{\added{Workflow of the experiments. $C_1i$ are selected (Latin square balanced) from the conditions \textit{baseline}~\base, \textit{highlight}~\highlight, and \textit{icon}~\icon, and $C_2i$ from the \textit{baseline 2}~\baseTwo and \textit{word-sized-graphics}~\wsgCond conditions.}}
	\label{fig:exp_setup}
	\Description{This Figure is a graphical representation of the timeline of the experiments.
		First is Experiment~1 with three conditions, then Experiment~2 with two conditions.}
\end{figure}
\added{Figure~\ref{fig:exp_setup} shows the temporal setup of both experiments.
	The independent variable is the experiment condition, and as dependent variables, we measured reading and question answering time, answer accuracy, AOI fixations, saccade amplitude, as well as qualitative ratings.
	We chose not to prime the participants, as we did not see any benefit and did not want to increase the complexity of the experiments.}

	\section{Results}

We investigate the results of both experiments with traditional performance analysis, i.e., correctness and completion times.
Further, we analyze differences in fixations and saccades between conditions. 
We evaluate significance based on the confidence intervals of pairwise differences~\cite{besancon_p_confidence_2017,cumming2013understanding,dragicevic2016fair}.
If those do not overlap with $0$, there is evidence of a difference.
This can be seen as equivalent to using p-value tests~\cite{krzywinski2013error}.
The further away the interval is from $0$, the stronger the evidence.
In this paper's OSF repository~\cite{osf}, there is data from the experiments, additional figures, all underlying data of the figures, as well as the scripts we used.
We further performed statistical inference on the data, details can be found in the OSF supplemental material.

\subsection{Performance}
\begin{figure*}
	\centering
	\captionsetup[subfigure]{labelformat=empty,singlelinecheck=false,justification=centering,aboveskip=-2pt,belowskip=-2pt}
	\hfill
	\begin{subfigure}[b]{.32\linewidth}
		\centering
		\includegraphics[width=.49\linewidth]{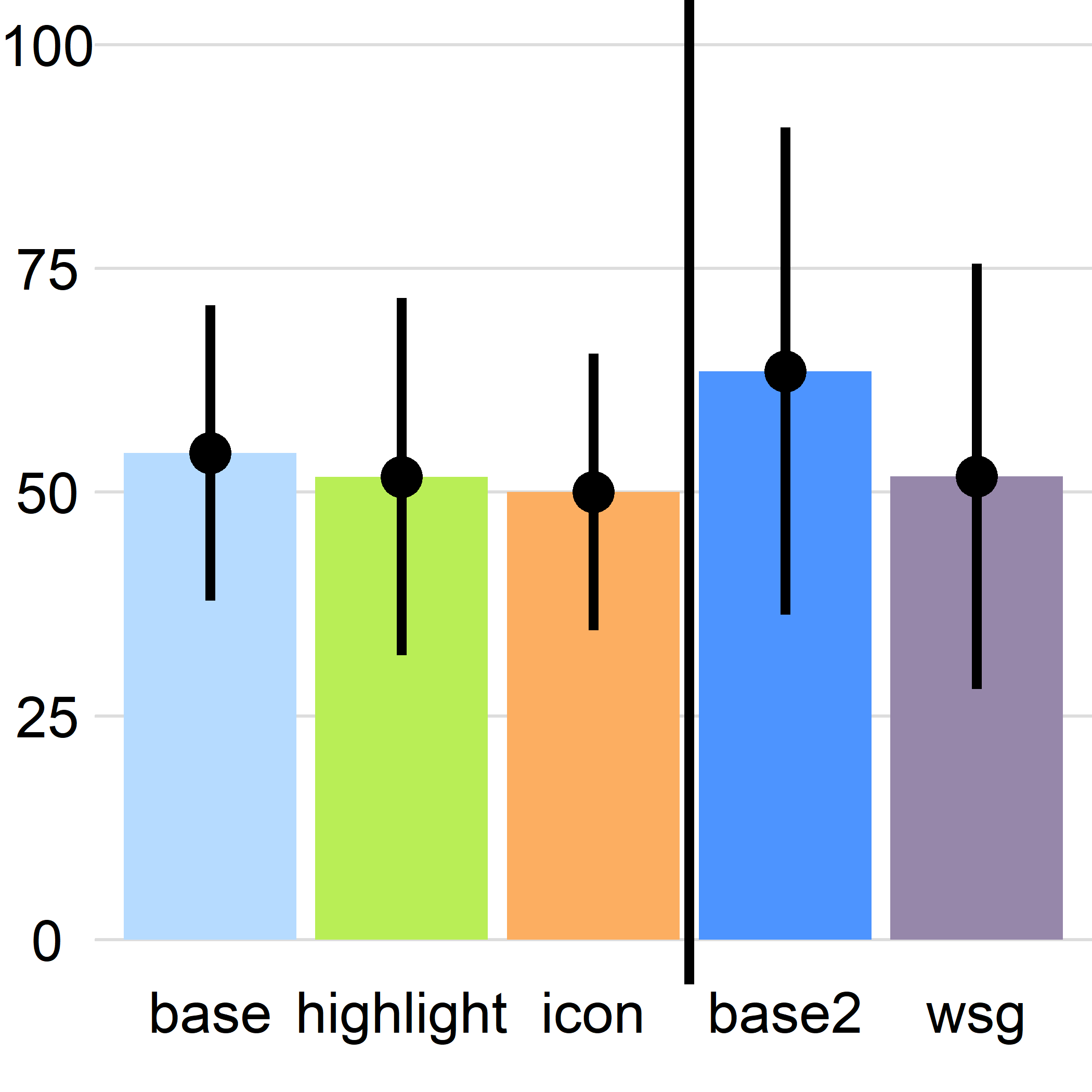}
		\begin{subfigure}[b]{.49\linewidth}
			\centering
			\includegraphics[width=\linewidth]{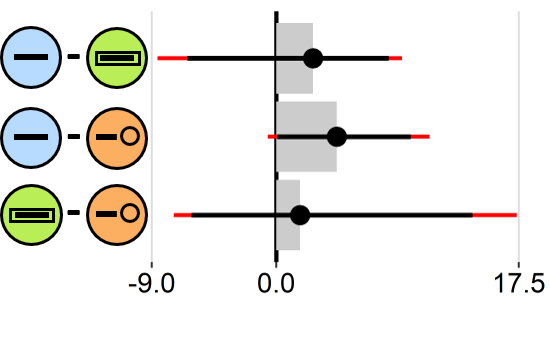}
			\includegraphics[width=\linewidth]{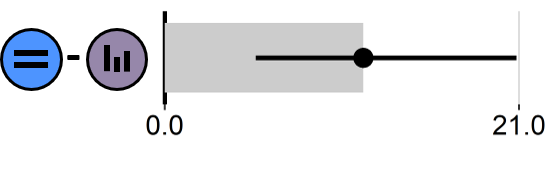}
		\end{subfigure}
		\caption{reading time (s)}
	\end{subfigure}
	\hfill
	\begin{subfigure}[b]{.32\linewidth}
		\centering
		\includegraphics[width=.49\linewidth]{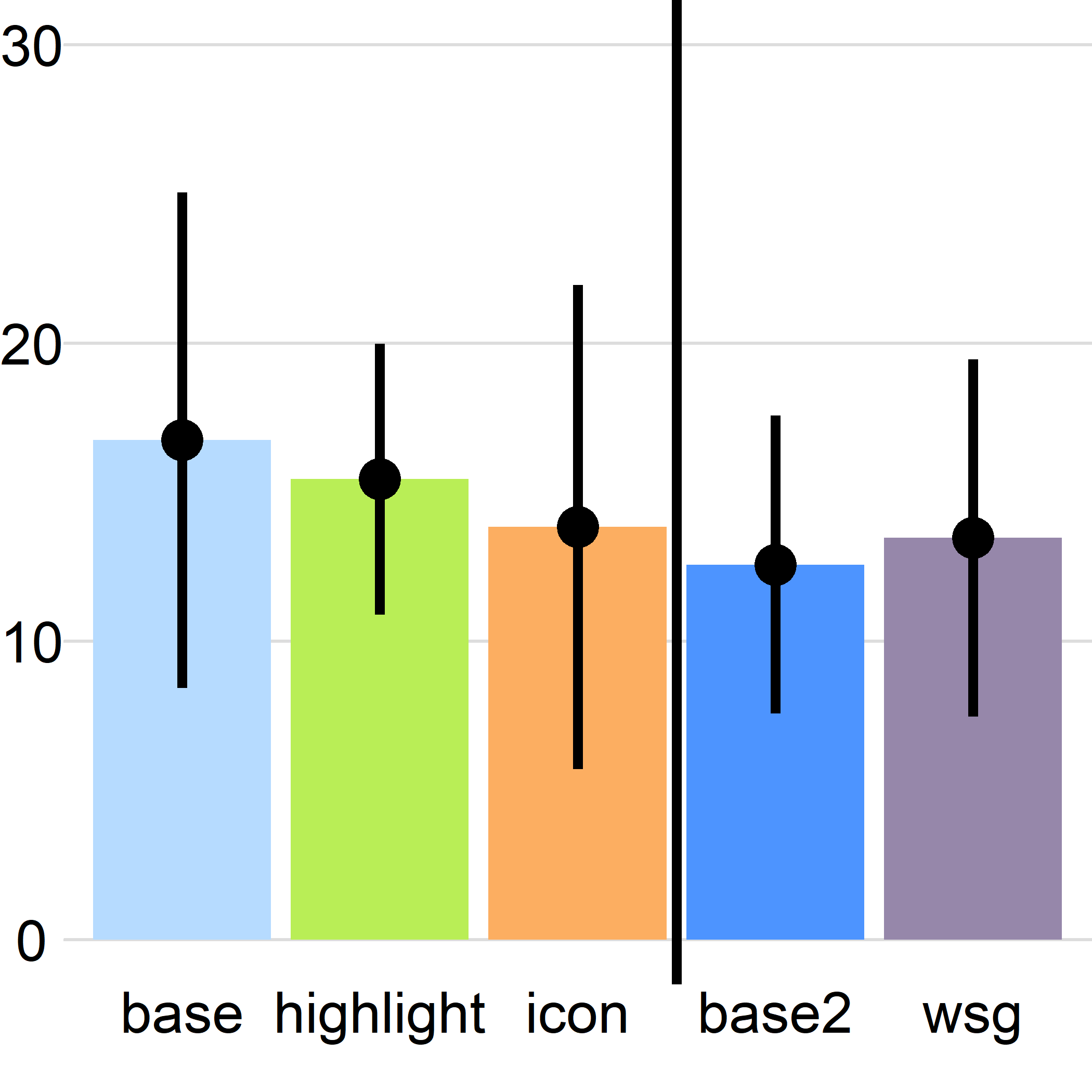}
		\begin{subfigure}[b]{.49\linewidth}
			\centering
			\includegraphics[width=\linewidth]{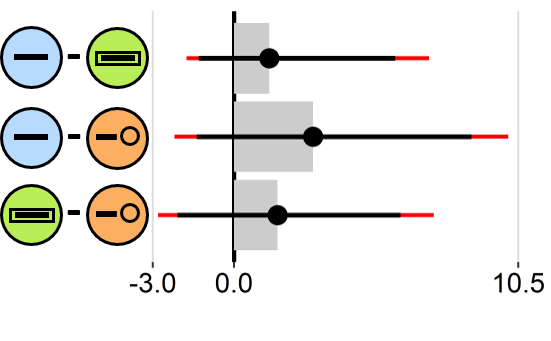}
			\includegraphics[width=\linewidth]{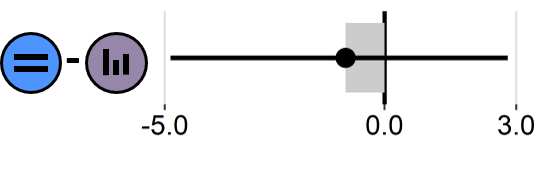}
		\end{subfigure}
		\caption{question answering time (s)}
	\end{subfigure}
	\hfill
	\begin{subfigure}[b]{.32\linewidth}
		\centering
		\includegraphics[width=.49\linewidth]{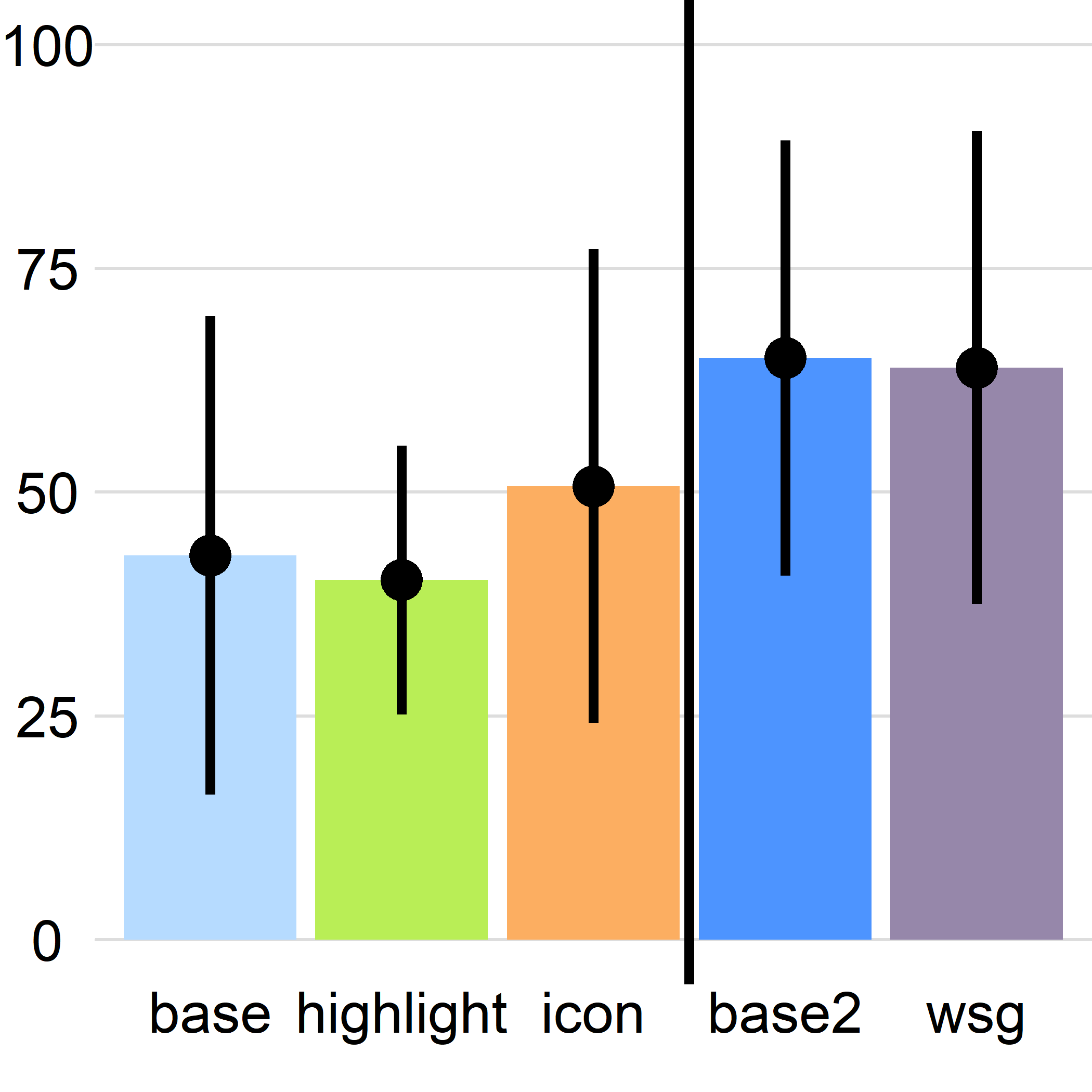}
		\begin{subfigure}[b]{.49\linewidth}
			\centering
			\includegraphics[width=\linewidth]{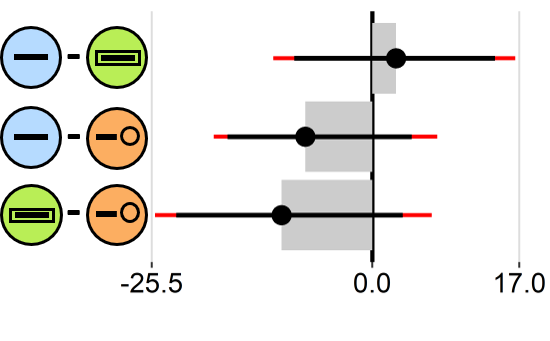}
			\includegraphics[width=\linewidth]{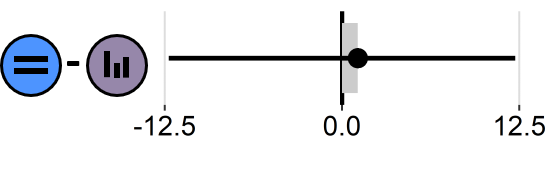}
		\end{subfigure}
		\caption{accuracy (\%)}
	\end{subfigure}
	\hfill
	\caption{\textit{Reading time} in seconds per stimulus, the time it took participants to answer the question for each stimulus, and answer accuracy in percent. Average with standard deviations (left chart, respectively). Pairwise differences between the study conditions (right chart, respectively), with 95\% bootstrap confidence intervals and red lines indicating the Bonferroni corrections for 3 pairwise comparisons in the first experiment.}
	\Description{
		Describes statistical analysis (means, CIs, pairwise differences).
		Reading time Experiment 1: For the baseline condition, the reading time has a mean of 54.37s and a CI of [45.25s, 63.3s], the highlight condition has a mean of 51.72s and a CI of [42.43s, 64.49s], and the icon condition has a mean of 50s and a CI of [42.3s, 59.41s].
		The mean of pairwise difference between the baseline and the highlight condition is 2.65, with a CI of [-8.58, 9.08].
		The mean of pairwise difference between the baseline and the icon condition is 4.36, with a CI of [-0.61, 11.07].
		The mean of pairwise difference between the highlight and the icon condition is 1.71, with a CI of [-7.4, 17.37].
		Reading time Experiment 2: For the baseline 2 condition, the reading time has a mean of 63.53s and a CI of [50.89s, 81.39s], and the word-sized graphics (wsg) condition has a mean of 51.77s and a CI of [40.82s, 67.13s].
		The mean of pairwise difference between the baseline 2 and the wsg condition is 11.76, with a CI of [5.39, 20.85].
		Question answering time Experiment 1: For the baseline condition, the time to answer the question has a mean of 16.75s and a CI of [13.10s, 22.28s], the highlight condition has a mean of 15.44s and a CI of [13.32s, 18.35s], and the icon condition has a mean of 13.83s and a CI of [10.78s, 20.68s].
		The mean of pairwise difference between the baseline and the highlight condition is 1.3, with a CI of [-1.75, 7.2].
		The mean of pairwise difference between the baseline and the icon condition is 2.91, with a CI of [-2.2, 10.13].
		The mean of pairwise difference between the highlight and the icon condition is 1.61, with a CI of [-2.81, 7.37].
		Question answering time Experiment 2: For the baseline 2 condition, the time to answer the question has a mean of 12.58s and a CI of [10.51s, 16.25s], and the word-sized graphics (wsg) condition has a mean of 13.47s and a CI of [10.56s, 17.05s].
		The mean of pairwise difference between the baseline 2 and the wsg condition is -0.89, with a CI of [-4.87, 2.81].
		Accuracy Experiment 1: For the baseline condition, the accuracy has a mean of 42.95\% and a CI of [28.85\%, 58.33\%], the highlight condition has a mean of 40.2\% and a CI of [32.84\%, 49.51\%], and the icon condition has a mean of 50.7\% and a CI of [34.03\%, 63.19\%].
		The mean of pairwise difference between the baseline and the highlight condition is 2.75, with a CI of [-11.43, 16.55].
		The mean of pairwise difference between the baseline and the icon condition is -7.75, with a CI of [-18.31, 7.53].
		The mean of pairwise difference between the highlight and the icon condition is -10.5, with a CI of [-25.12, 6.88].
		Accuracy Experiment 2: For the baseline 2 condition, the accuracy has a mean of 65\% and a CI of [50\%, 76.67\%], and the word-sized graphics (wsg) condition has a mean of 63.89\% and a CI of [47.22\%, 75\%].
		The mean of pairwise difference between the baseline 2 and the wsg condition is 1.11, with a CI of [-12.22, 12.22].
	}
	\label{fig:time_and_accuracy}
\end{figure*}

Figure~\ref{fig:time_and_accuracy} shows the results for text reading time, question answering time, and accuracy.
Since time was not limited, a relative increase in task duration between conditions for each participant might be an indicator that they had problems remembering the stimulus.

\paragraph{Experiment 1}
Participants took a similar amount of time to read the stimulus texts for each of the conditions in Experiment~1, between~$50$ and~$55$~seconds on average~(Figure~\ref{fig:time_and_accuracy}, left).
The reading time is slightly higher in the \textit{baseline}~\base condition, and lowest for text with \textit{icons}~\icon, but there are no significant differences.
The average question answering time~(Figure~\ref{fig:time_and_accuracy}, center) shows a similar pattern.
Again, participants took the longest for the \textit{baseline}~\base condition, and least for the \textit{icon}~\icon condition, but without significant differences.
The accuracy of the responses~(Figure~\ref{fig:time_and_accuracy}, right) is highest for texts with \textit{icons}~\icon at around~$50\%$ on average, lowest for texts with \textit{highlights}~\highlight at~$40\%$, with no significant differences.

\paragraph{Experiment 2}
The reading time for each stimulus text~(Figure~\ref{fig:time_and_accuracy}, left) in Experiment~2 is significantly higher in the \textit{baseline 2}~\baseTwo condition~($64$ seconds on average) than for texts with \textit{word-sized graphics}~(\textit{wsg}~\wsgCond, $52$ seconds).
In contrast, question answering time~(Figure~\ref{fig:time_and_accuracy}, center) is slightly higher in the \textit{wsg}~\wsgCond~condition.
The accuracy~(Figure~\ref{fig:time_and_accuracy}, right) is similar for both conditions in Experiment~2 at around~$65\%$.
Overall, we found no significant differences in answering time and accuracy.

\subsection{Focus on Visual Elements}
\begin{figure}[b]
	\centering
	\includegraphics[width=\linewidth]{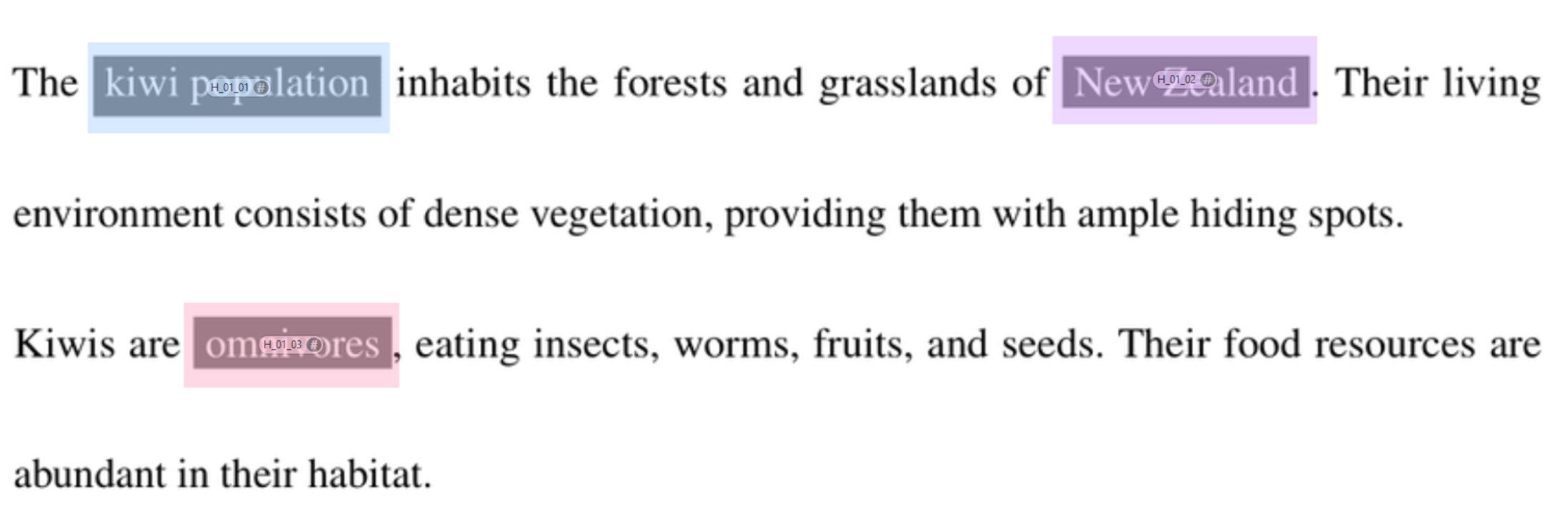}
	\hfill
	\hrule
	\vspace{1em}
	\includegraphics[width=\linewidth]{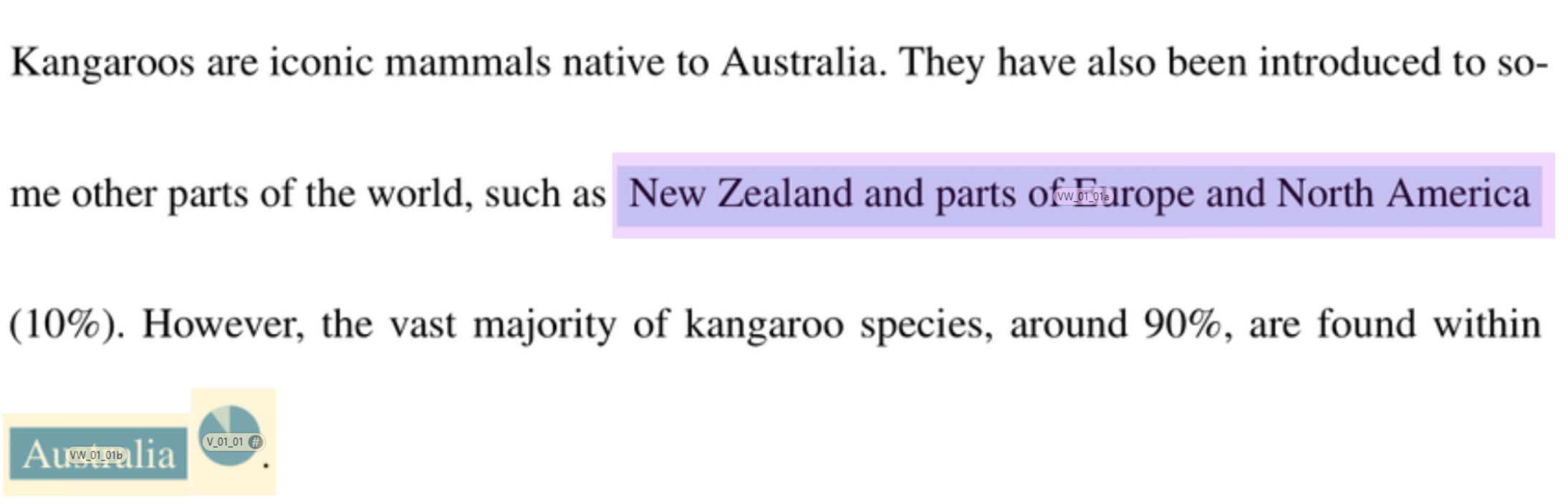}
	\caption{Example of the AOIs we defined for stimuli for the experiment conditions (top)~\textit{highlight}~\highlight, and (bottom)~\textit{word-sized graphics}~\wsgCond.}
	\Description{This figure shows two images of text about animals, with some of the factual terms in the text highlighted with a colored background.
		At the end of the sentence that has the colored terms, there is a small chart that depicts the percentages of the facts.
		Additionally, the highlights and graphics are marked as areas of interest (AOIs), while the rest of the text is not.}
	\label{fig:stimuli_aoi}
\end{figure}
We defined the visualization elements as individual areas of interest (AOIs) to investigate transitions from text reading to visual elements.
Examples of these AOIs can be found in Figure~\ref{fig:stimuli_aoi}.
For the \textit{baseline}~\base~\baseTwo conditions, we defined AOIs on facts that are equivalent to the ones that were augmented with visual elements in the other conditions.
We investigated the relative duration of fixations on AOIs to identify potential distraction and the average saccade amplitude as indicator for jumps in reading in addition to the typical line regressions.

\begin{figure*}[t]
	\centering
	\captionsetup[subfigure]{labelformat=empty,singlelinecheck=false,justification=centering,aboveskip=-2pt,belowskip=-2pt}
	\hfill
	\begin{subfigure}[b]{.47\linewidth}
		\centering
		\includegraphics[width=.49\linewidth]{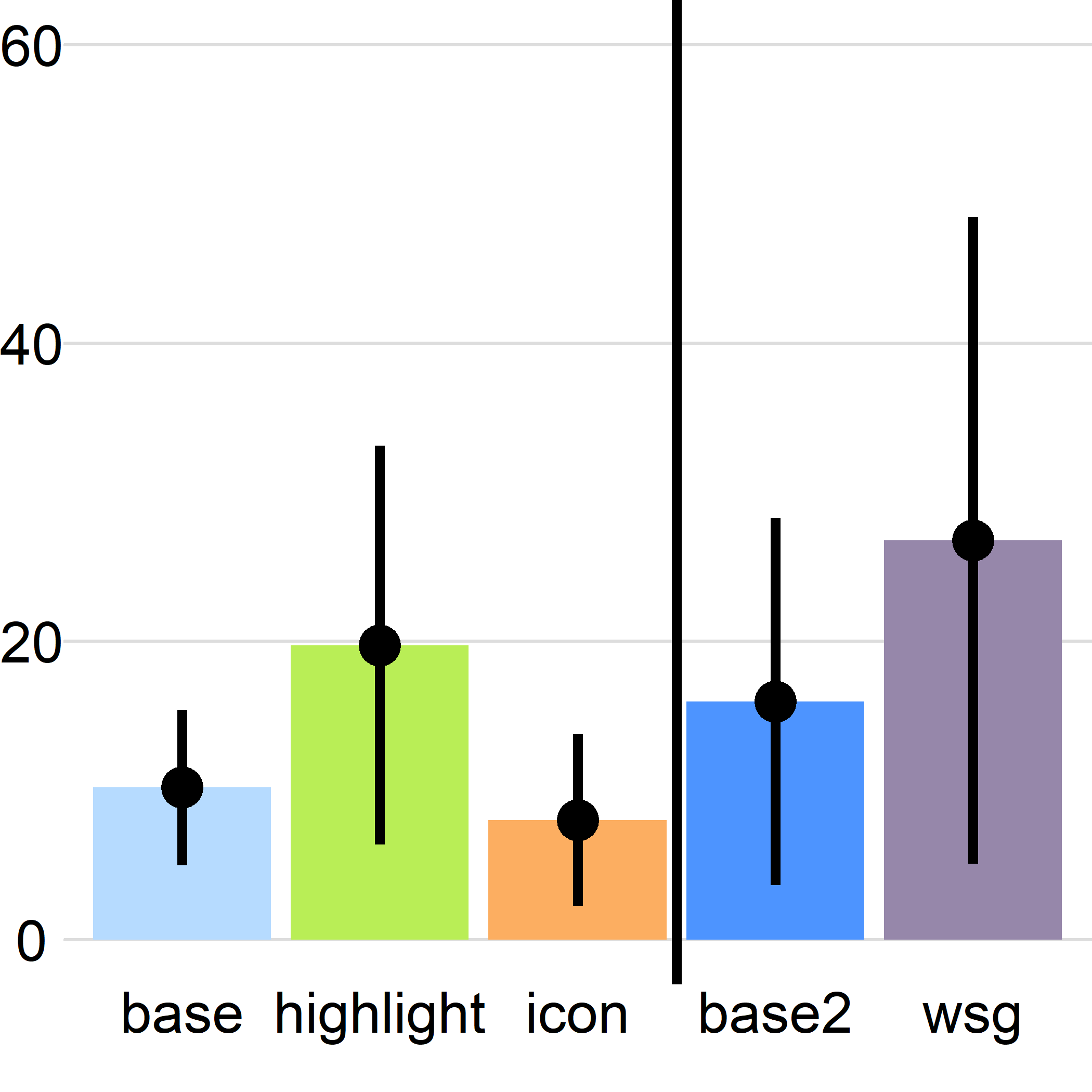}
		\begin{subfigure}[b]{.49\linewidth}
			\centering
			\includegraphics[width=\linewidth]{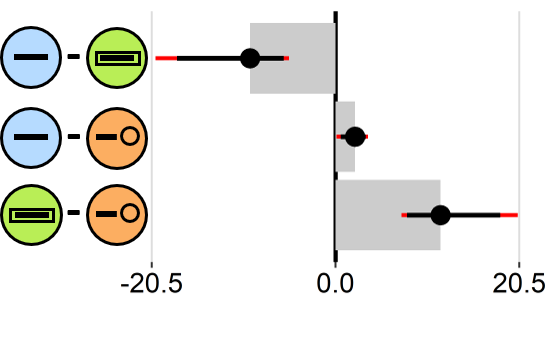}
			\includegraphics[width=\linewidth]{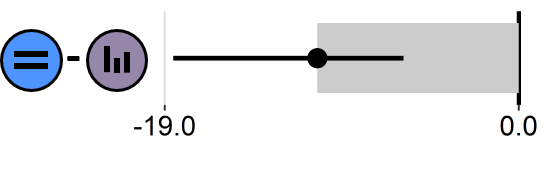}
		\end{subfigure}
		\caption{fixation on AOIs (\%)}
	\end{subfigure}
	\hfill
	\begin{subfigure}[b]{.47\linewidth}
		\centering
		\includegraphics[width=.49\linewidth]
		{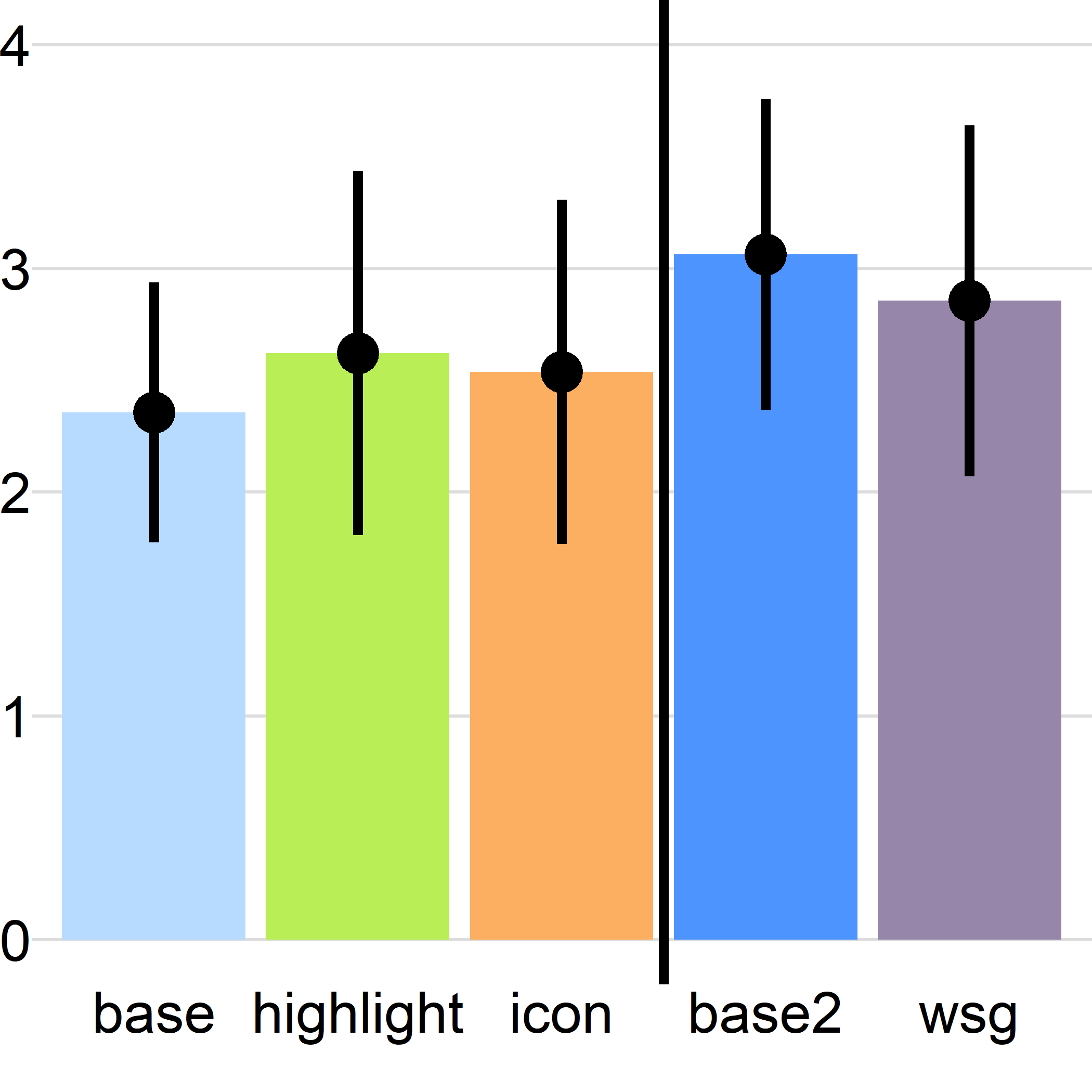}
		\begin{subfigure}[b]{.49\linewidth}
			\centering
			\includegraphics[width=\linewidth]{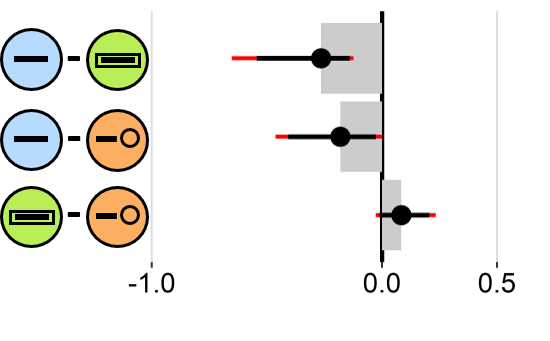}
			\includegraphics[width=\linewidth]{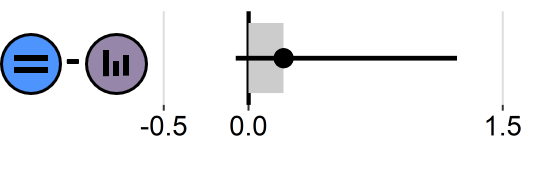}
		\end{subfigure}
		\caption{average saccade amplitude (°)}
	\end{subfigure}
	\hfill
	\caption{\textit{Duration of fixations} on visual elements (and for the baseline conditions, equivalent facts in the texts) in relation to the reading time per stimulus, and the average amplitude of saccades. Average with standard deviations (left chart, respectively). Pairwise differences between the study conditions (right chart, respectively), with 95\% bootstrap confidence intervals and red lines indicating the Bonferroni corrections for 3 pairwise comparisons in the first experiment.}
	\Description{
		Describes statistical analysis (means, CIs, pairwise differences).
		Fixations on AOIs Experiment 1: For the baseline condition, the fixation duration on AOIs has a mean of 10.21\% and a CI of [7.53\%, 13.22\%], the highlight condition has a mean of 19.75\% and a CI of [14.57\%, 31.27\%], and the icon condition has a mean of 8.03\% and a CI of [5.63\%, 12.39\%].
		The mean of pairwise difference between the baseline and the highlight condition is -9.54, with a CI of [-20.08, -5.18].
		The mean of pairwise difference between the baseline and the icon condition is 2.18, with a CI of [0.13, 3.63].
		The mean of pairwise difference between the highlight and the icon condition is 11.72, with a CI of [7.37, 20.36].
		Fixations on AOIs Experiment 2: For the baseline 2 condition, the fixation duration on AOIs has a mean of 16\% and a CI of [10.58\%, 24.63\%], and the word-sized graphics (wsg) condition has a mean of 26.8\% and a CI of [17.83\%, 43.26\%].
		The mean of pairwise difference between the baseline 2 and the wsg condition is -10.81, with a CI of [-18.55, -6.19].
		Saccade amplitude Experiment 1: For the baseline condition, the average saccade amplitude has a mean of 2.36\% and a CI of [1.98\%, 2.63\%], the highlight condition has a mean of 2.62\% and a CI of [2.17\%, 3.05\%], and the icon condition has a mean of 2.54\% and a CI of [2.10\%, 2.94\%].
		The mean of pairwise difference between the baseline and the highlight condition is -0.27, with a CI of [-0.65, -0.12].
		The mean of pairwise difference between the baseline and the icon condition is -0.18, with a CI of [-0.46, 0.001].
		The mean of pairwise difference between the highlight and the icon condition is 0.08, with a CI of [-0.03, 0.23].
		Saccade amplitude Experiment 2: For the baseline 2 condition, the average saccade amplitude has a mean of 3.06\% and a CI of [2.73\%, 3.47\%], and the word-sized graphics (wsg) condition has a mean of 2.86\% and a CI of [2.38\%, 3.24\%].
		The mean of pairwise difference between the baseline 2 and the wsg condition is 0.21, with a CI of [-0.08, 1.23].
	}
	\label{fig:fixations_and_saccades}
\end{figure*}


\paragraph{Experiment 1}
In relation to the reading time, the percentage of fixations on visual elements is significantly higher for \textit{highlights}~\highlight than for \textit{icons}~\icon and plain text~\base~(Figure~\ref{fig:fixations_and_saccades}, left).
The duration of fixations on visual elements (or factual terms) is also significantly higher for plain texts~\base than for texts with \textit{icons}~\icon.
Regarding saccades~(Figure~\ref{fig:fixations_and_saccades}, right), the average amplitude was similar for the \textit{highlight}~\highlight and \textit{icon}~\icon conditions, but significantly lower between the \textit{baseline}~\base and \textit{highlights}~\highlight. 

\paragraph{Experiment 2}
In the second experiment (Figure~\ref{fig:fixations_and_saccades}), the average relative fixation duration on AOIs was significantly higher with visual enhancements~\wsgCond than for the \textit{baseline}~\baseTwo.
The saccade amplitude was slightly higher in the \textit{baseline}~\baseTwo condition than for \textit{word-sized graphics}~\wsgCond with no significant difference.

\subsection{Subjective Feedback}
\begin{figure*}
	\centering
	\captionsetup[subfigure]{labelformat=empty,aboveskip=-2pt,belowskip=-2pt}
	\begin{subfigure}[b]{.47\linewidth}
		\centering
		\includegraphics[width=.48\linewidth]{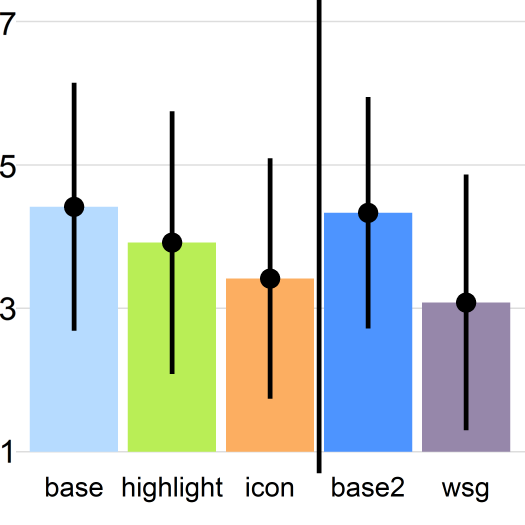}
		\begin{subfigure}[b]{.48\linewidth}
			\centering
			\includegraphics[width=\linewidth]{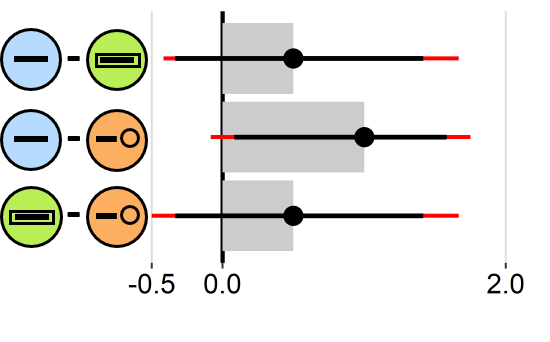}
			\includegraphics[width=\linewidth]{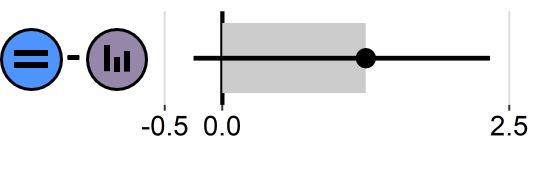}
		\end{subfigure}
		\caption{difficulty}
	\end{subfigure}
	\hfill
	\begin{subfigure}[b]{.47\linewidth}
		\centering
		\includegraphics[width=.48\linewidth]{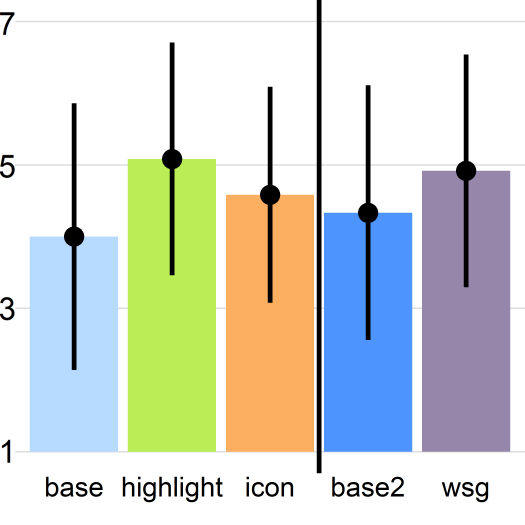}
		\begin{subfigure}[b]{.48\linewidth}
			\centering
			\includegraphics[width=\linewidth]{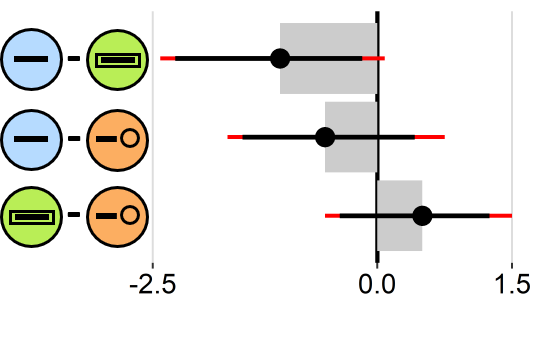}
			\includegraphics[width=\linewidth]{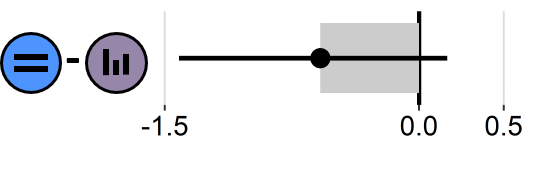}
		\end{subfigure}
		\caption{confidence}
	\end{subfigure}
	\hfill
	\begin{subfigure}[b]{.47\linewidth}
		\centering
		\includegraphics[width=.48\linewidth]{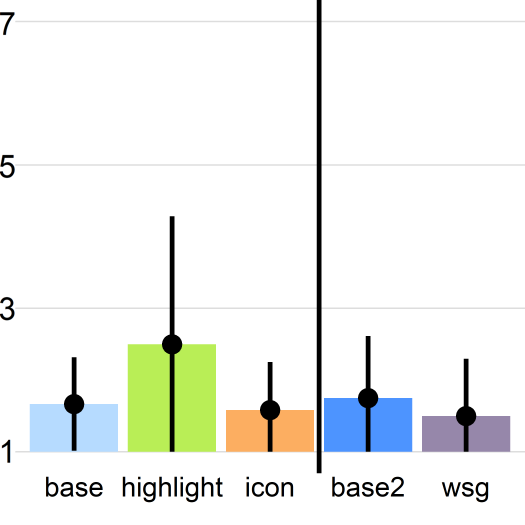}
		\begin{subfigure}[b]{.48\linewidth}
			\centering
			\includegraphics[width=\linewidth]{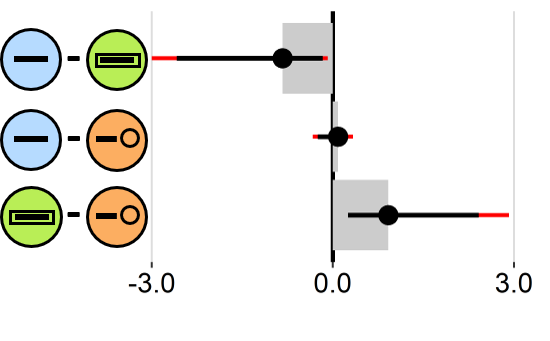}
			\includegraphics[width=\linewidth]{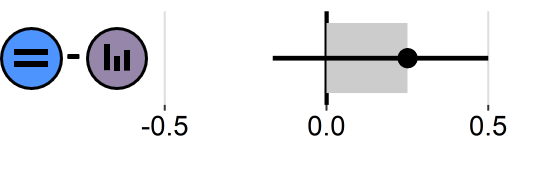}
		\end{subfigure}
		\caption{rushed}
	\end{subfigure}
	\hfill
	\begin{subfigure}[b]{.47\linewidth}
		\centering
		\includegraphics[width=.48\linewidth]{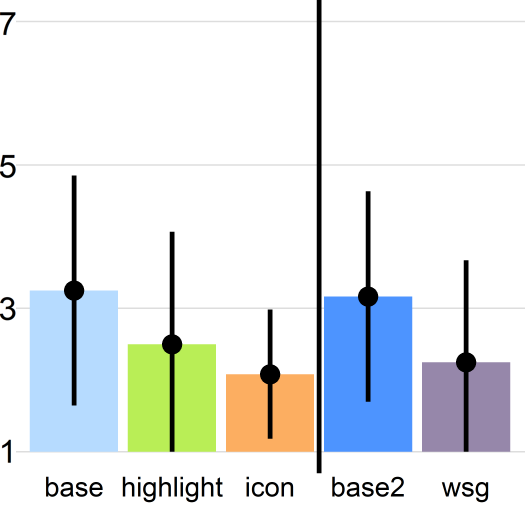}
		\begin{subfigure}[b]{.48\linewidth}
			\centering
			\includegraphics[width=\linewidth]{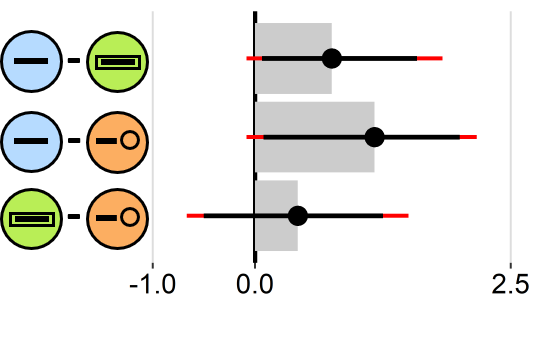}
			\includegraphics[width=\linewidth]{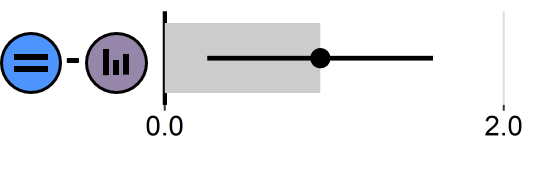}
		\end{subfigure}
		\caption{annoyed}
	\end{subfigure}
	\caption{Qualitative ratings of participants' perceived \textit{task difficulty}, \textit{confidence} in answers, how \textit{rushed} and how \textit{annoyed} they felt, from low (1) to high (7).
		Average with standard deviations (left chart, respectively). Pairwise differences between the study conditions (right chart, respectively), with 95\% bootstrap confidence intervals and red lines indicating the Bonferroni corrections for 3 pairwise comparisons in the first experiment.}
	\label{fig:ratings}
	\Description{
		Describes statistical analysis (means, CIs, pairwise differences).
		Difficulty Experiment 1: For the baseline condition, the difficulty has a mean rating of 3.42 and a CI of [2.25, 4.17], the highlight condition has a mean rating of 2.92 and a CI of [1.83, 3.83], and the icon condition has a mean rating of 2.42 and a CI of [1.58, 3.42].
		The mean of pairwise difference between the baseline and the highlight condition is 0.5, with a CI of [-0.42, 1.67].
		The mean of pairwise difference between the baseline and the icon condition is 1, with a CI of [-0.08, 1.75].
		The mean of pairwise difference between the highlight and the icon condition is 0.5, with a CI of [-0.5, 1.67].
		Difficulty Experiment 2: For the baseline 2 condition, the difficulty has a mean rating of 3.3 and a CI of [2.17, 4], and the word-sized graphics (wsg) condition has a mean rating of 2.08 and a CI of [1.25, 3.17].
		The mean of pairwise difference between the baseline 2 and the wsg condition is 1.25, with a CI of [-0.25, 2.33].
		Confidence Experiment 1: For the baseline condition, the confidence has a mean rating of 3 and a CI of [1.92, 3.92], the highlight condition has a mean rating of 4.08 and a CI of [2.83, 4.75], and the icon condition has a mean rating of 3.58 and a CI of [2.67, 4.33].
		The mean of pairwise difference between the baseline and the highlight condition is -1.08, with a CI of [-2.42, 0.08].
		The mean of pairwise difference between the baseline and the icon condition is -0.58, with a CI of [-1.67, 0.75].
		The mean of pairwise difference between the highlight and the icon condition is 0.5, with a CI of [-0.58, 1.5].
		Confidence Experiment 2: For the baseline 2 condition, the confidence has a mean rating of 3.33 and a CI of [2.17, 4.17], and the word-sized graphics (wsg) condition has a mean rating of 3.92 and a CI of [2.67, 4.58].
		The mean of pairwise difference between the baseline 2 and the wsg condition is -0.58, with a CI of [-1.42, 0.17].
		Rushed Experiment 1: For the baseline condition, rushed has a mean rating of 0.67 and a CI of [0.25, 1], the highlight condition has a mean rating of 1.5 and a CI of [0.75, 2.75], and the icon condition has a mean rating of 0.58 and a CI of [0.25, 0.92].
		The mean of pairwise difference between the baseline and the highlight condition is -0.83, with a CI of [-3, -0.08].
		The mean of pairwise difference between the baseline and the icon condition is 0.08, with a CI of [-0.33, 0.33].
		The mean of pairwise difference between the highlight and the icon condition is 0.92, with a CI of [0.25, 2.92].
		Rushed Experiment 2: For the baseline 2 condition, rushed has a mean rating of 0.75 and a CI of [0.33, 1.33], and the word-sized graphics (wsg) condition has a mean rating of 0.5 and a CI of [0.08, 0.92].
		The mean of pairwise difference between the baseline 2 and the wsg condition is 0.25, with a CI of [-0.17, 0.5].
		Annoyed Experiment 1: For the baseline condition, annoyed has a mean rating of 2.25 and a CI of [1.25, 3], the highlight condition has a mean rating of 1.5 and a CI of [0.75, 2.42], and the icon condition has a mean rating of 1.08 and a CI of [0.58, 1.5].
		The mean of pairwise difference between the baseline and the highlight condition is 0.75, with a CI of [-0.08, 1.83].
		The mean of pairwise difference between the baseline and the icon condition is 1.17, with a CI of [-0.08, 2.17].
		The mean of pairwise difference between the highlight and the icon condition is 0.42, with a CI of [-0.67, 1.5].
		Annoyed Experiment 2: For the baseline 2 condition, annoyed has a mean rating of 2.17 and a CI of [1.25, 2.83], and the word-sized graphics (wsg) condition has a mean rating of 1.25 and a CI of [0.58, 2.17].
		The mean of pairwise difference between the baseline 2 and the wsg condition is 0.92, with a CI of [0.25, 1.58].
	}
	\vspace{-.5em}
\end{figure*}

Further, we collected subjective feedback with respect to difficulty, confidence, and the impressions of being rushed and annoyed.
Figure~\ref{fig:ratings} shows the participants' ratings of each condition, \removed{They}\added{which} were given on a $1$~to~$7$~Likert scale\removed{, which we adjusted for the bar~charts to start at~$0$}.
The perceived difficulty is highest for the \textit{baseline}~\base~\baseTwo conditions in both experiments and in Experiment~1 lowest for texts with \textit{icons}~\icon.
The confidence in the answers was rated highest for texts with \textit{highlights}~\highlight in Experiment~1, and for texts with \textit{wsg}~\wsgCond in Experiment~2.
For the \textit{baseline}~\base~\baseTwo conditions, the confidence rating was lowest in both experiments.
These differences are, however, not significant.

Participants rated that they felt rushed or hurried significantly more for texts with \textit{highlights}~\highlight than for the other conditions in Experiment~1.
In Experiment~2, participants felt slightly more rushed in the \textit{baseline}~\baseTwo than the \textit{wsg}~\wsgCond~condition.
They also felt significantly more annoyed or stressed in the \textit{baseline}~\baseTwo than \textit{wsg}~\wsgCond~condition.
In Experiment~1, participants reported most annoyance with the \textit{baseline}~\base condition and least with texts with \textit{icons}~\icon, but without a significant difference.

\paragraph{Experiment condition preferences}
Almost all participants reported that they preferred texts with visual elements~\highlight~\icon~\wsgCond over plain texts~\base~\baseTwo, in both experiments, as they ``make it easier to identify relevant parts'' and are ``helpful for memorization''.
Two participants found the \textit{highlighted}~\highlight texts or the texts with \textit{word-sized graphics}~\wsgCond too distracting, but still liked \textit{icons}~\icon more than plain text~\base.
In Experiment~1, about half of the participants preferred \textit{icons}~\icon over \textit{highlights}~\highlight, while the others liked \textit{highlights}~\highlight more.
In Experiment~2, participants especially liked the bar charts best and found them easiest to read.

\paragraph{Reading strategies}
Most participants claimed that they usually do not follow a specific reading strategy.
Some said that they sometimes skim over filler words, or read quickly and not word-for-word.
For the baseline~\base~\baseTwo conditions, some participants reported reading the text and then going through it again and ``trying to remember possible key information'', and ``after a few stimuli, started focusing more on the fact lists in the text''.
With visual elements~\highlight~\icon~\wsgCond, most participants' self-reported reading strategies changed.
They said they focused more on the visual elements than on the rest of the text, and, after reading the whole text, went back to the visual elements again to help with memorizing information.
For example, they ``tried to link the word with the icon''~\icon, or were ``skimming over it {[}the text{]} and focused on the highlights/icons''~\highlight~\icon.
In Experiment~2, some participants stated that they inspected the word-sized graphics~\wsgCond more closely after reading the text, to better memorize the facts.

\begin{figure*}
	\centering
	\captionsetup[subfigure]{labelformat=empty,singlelinecheck=false,justification=centering,aboveskip=-2pt,belowskip=-2pt}
	\hfill
	\begin{subfigure}{.47\linewidth}
		\includegraphics[width=\linewidth]{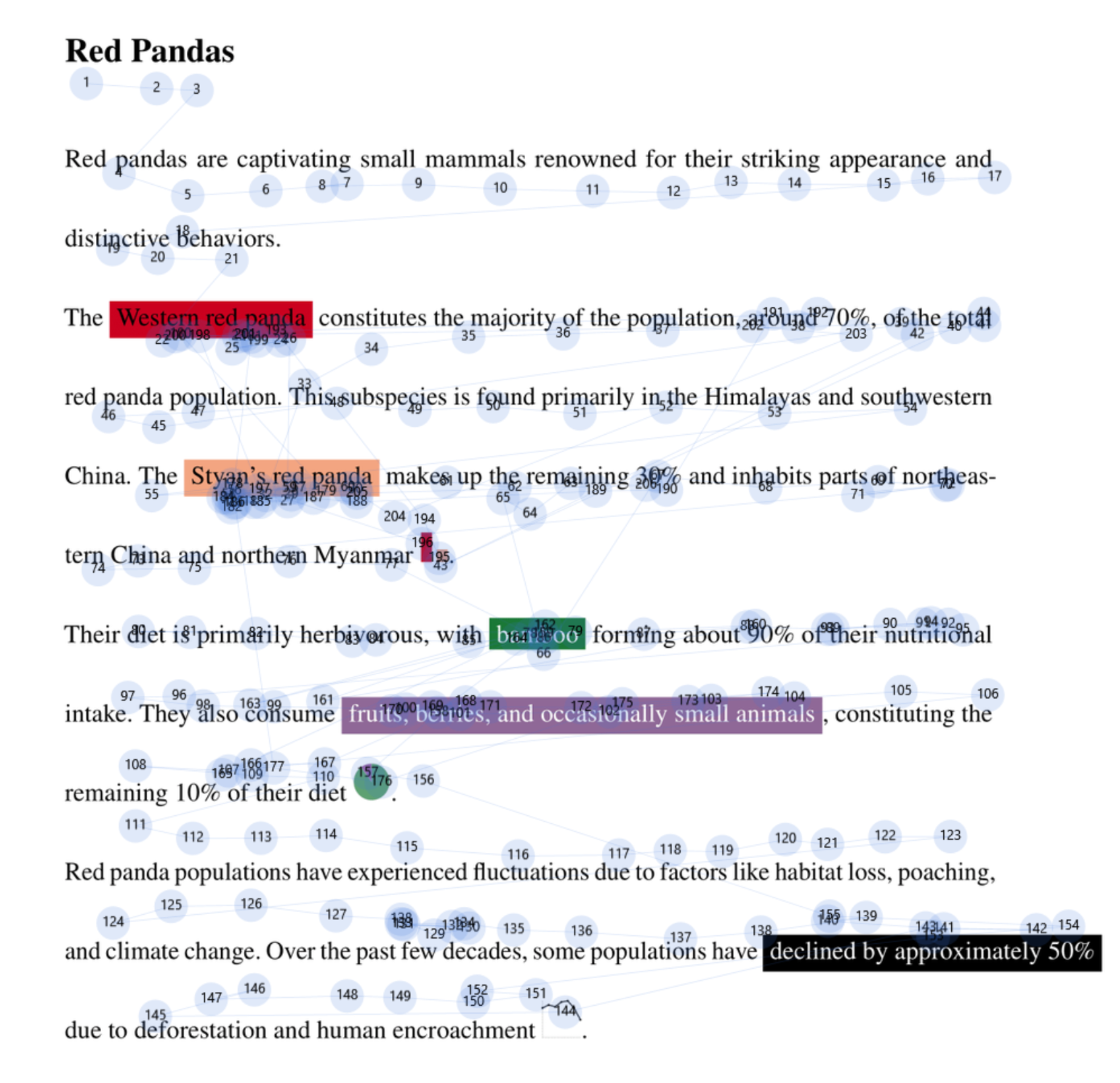}
	\end{subfigure}
	\hfill
	\begin{subfigure}{.47\linewidth}
		\includegraphics[width=\linewidth]{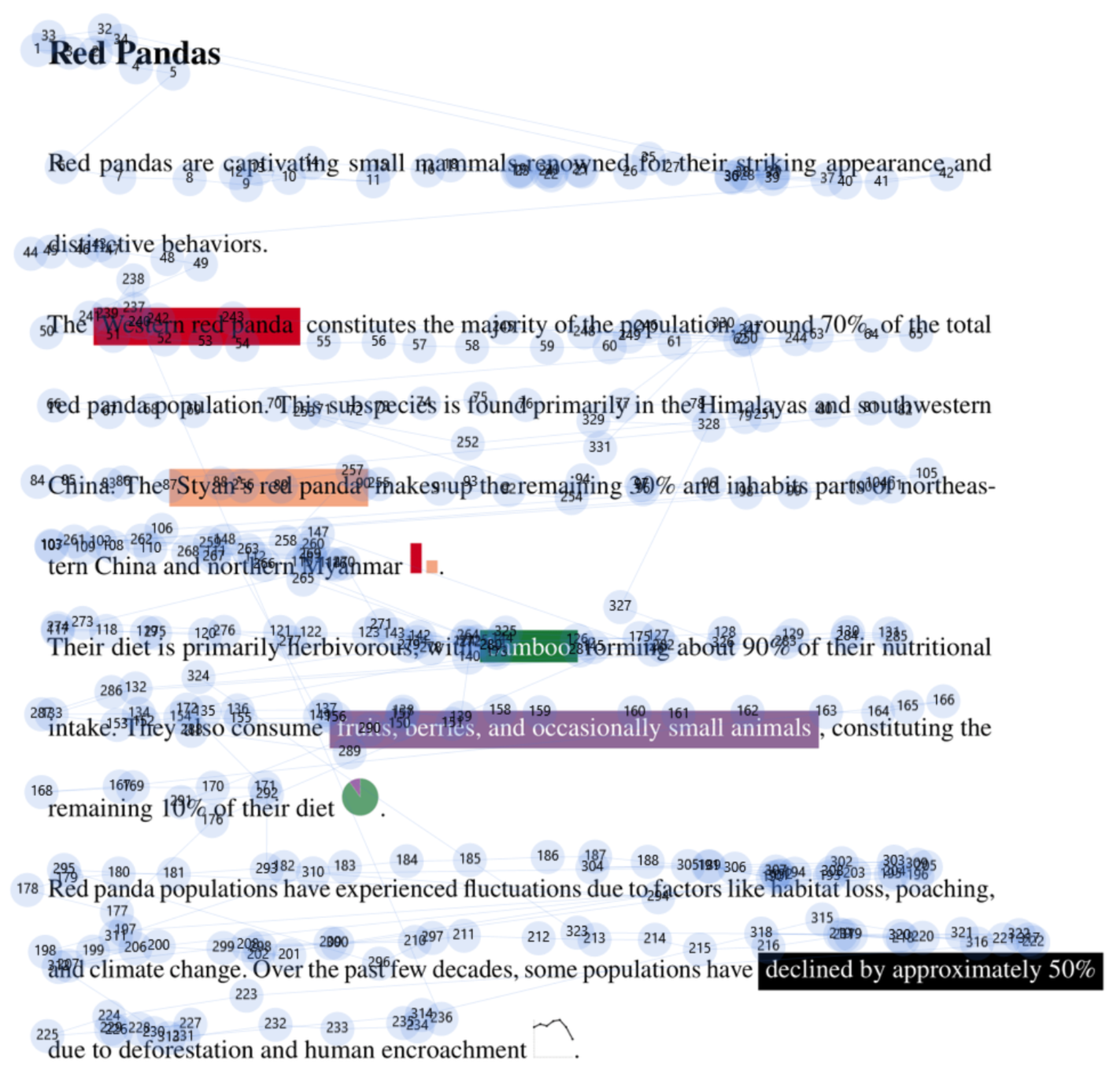}
	\end{subfigure}
	\hfill
	\caption{Gaze Plots of participants P4 (left) and P9 (right) on the \textit{word-sized graphics}~\wsgCond condition. Participants received the following question: \emph{How much of the total red panda population does the Western red panda make up?} P4 read the highlighted parts frequently and answered the question correctly, whereas P9 did not.}
	\Description{This figure shows two images of text about red pandas, with some of the factual terms in the text highlighted with a colored background.
		At the end of the sentence that has the colored terms, there is a small chart that depicts the percentages of the facts.
		Overlayed on the text is a gaze plot, i.e., numbered circled connected with lines to show where someone looked, and in which order, when reading the text.
		On the left, the image shows a gaze plot of a participant with correct answer to a question about the text.
		The circles that depict the fixations are mostly on the highlighted parts of the text.
		On the right, the gaze plot of a participant with a wrong answer to the question about the text shows fixations on the highlights, but also a lot of other parts of the text.}
	\label{fig:heatmap_example}
\end{figure*}

\subsection{Correlations}
As is to be expected, we found an inverse correlation between the difficulty and confidence ratings.
There is also a correlation between the ratings for difficulty and level of annoyance.
There are, however, no remarkable correlations between task performance and any other factors.
We included further details in our supplemental material.

\section{Discussion}
Our preliminary results indicate that there are differences in reading behavior when text is enhanced with additional information.
Some results confirm common expectations from experience and from the literature.
Others, such as differences in results between the \textit{highlight} and \textit{icon} conditions, are more subtle. 

\paragraph{Visual Elements: Distraction or help?}
Our results showed that especially highlights in text attract attention.
This result was expected; highlighted text has a pop-out effect, making it pre-attentively perceivable and looked at more frequently than words in plain text.
The main question in this case is: \textit{Is this good or bad?}
Highlighted texts (also with word-sized graphics) were found distracting by some participants, which is reflected in the higher fixation duration for \textit{highlight}~\highlight and \textit{word-sized graphics}~\wsgCond than for \textit{icon}~\icon.
Text that was augmented with \textit{icons}~\icon overall received the best ratings, and participants tended to perform best. 
The baseline condition~\base shows longer fixations than \textit{icon}~\icon, which could indicate that with icons it was easier to grasp the factual terms, while not being as distracting as highlights~\highlight.
We further noticed frequent saccades to the visual elements~\highlight~\icon~\wsgCond, which is also reflected by the lower saccade amplitudes for plain text~\base~\baseTwo.
This could be one reason for some people feeling distracted, but it might also help them remember these particular elements better (Figure~\ref{fig:heatmap_example}, left).
Further experiments will be necessary to confirm these assumptions.

\paragraph{Hypotheses for further studies}
We summarized our results in the form of hypotheses we plan to evaluate in future studies that focus less on the exploratory aspects of text enhancements and more on hypothesis-driven testing with specified tasks and stimuli.

\begin{itemize}
    \item [\textbf{H$_\text{1}$:}] The fixations for texts augmented with icons are lower than for plain texts.
    \item [\textbf{H$_\text{2}$:}] Saccade amplitudes for text with visual enhancements are higher due to more frequent looks on these elements.
    \item [\textbf{H$_\text{3}$:}] Information memorization and reading speed are higher for texts with visual elements than for plain texts.
\end{itemize}
Further, there is a subjective preference for texts with visual enhancements over plain texts which we also plan to quantify and investigate in more detail in the future.

	
\section{Conclusion}
We presented the results of a small-scale study investigating the influence of highlighting, symbols, and word-sized graphics on reading behavior for text documents.
To this point, the conducted experiments were exploratory to observe differences between conditions and build hypotheses.
\added{We performed the experiments with just $12$~participants to quickly achieve those first results.}
For future work, we plan to conduct a larger study with more statistical power to confirm our findings.

The presented results support findings from the literature~\cite{fowler1974effectiveness, lorch1989text, ben2018contrib, goffin_exploring_2015} that visual enhancements help remember facts of a text better.
By investigating the eye movements during reading, we could see that people tend to look at the added elements frequently, which can disturb a consistent reading flow and is perceived as annoying by some participants.
Hence, the use of visual enhancements, especially highlights and word-sized graphics, should be considered carefully, focusing on the most important elements without overloading the text with additional elements.
\added{Further, there are challenges for choosing suitable icons.
They should be clear in the terms and concepts they depict, as well as respect cultural differences.}

	\begin{acks}
		We thank all participants of the study.
		This research was funded by the Deutsche Forschungsgemeinschaft (DFG, German Research Foundation) – project 314647693 (VA4VGI) as part of the Priority Program \textit{VGIscience} (SPP 1894), by DFG project 449742818, and under Germany’s Excellence Strategy - EXC 2075 - 390740016.
		Additionally, we acknowledge the funding by the Deutsche Forschungsgemeinschaft (DFG, German Research Foundation) -- Project-ID 251654672 -- TRR 161 (Project B01).
		We thank Petra Isenberg for early discussions on the topic.
	\end{acks}
	
	\bibliographystyle{ACM-Reference-Format}
	\bibliography{refs.bib}


\begin{thebibliography}{29}


\ifx \showCODEN    \undefined \def \showCODEN     #1{\unskip}     \fi
\ifx \showDOI      \undefined \def \showDOI       #1{#1}\fi
\ifx \showISBNx    \undefined \def \showISBNx     #1{\unskip}     \fi
\ifx \showISBNxiii \undefined \def \showISBNxiii  #1{\unskip}     \fi
\ifx \showISSN     \undefined \def \showISSN      #1{\unskip}     \fi
\ifx \showLCCN     \undefined \def \showLCCN      #1{\unskip}     \fi
\ifx \shownote     \undefined \def \shownote      #1{#1}          \fi
\ifx \showarticletitle \undefined \def \showarticletitle #1{#1}   \fi
\ifx \showURL      \undefined \def \showURL       {\relax}        \fi
\providecommand\bibfield[2]{#2}
\providecommand\bibinfo[2]{#2}
\providecommand\natexlab[1]{#1}
\providecommand\showeprint[2][]{arXiv:#2}

\bibitem[Abdelaal et~al\mbox{.}(2023)]%
        {9908291}
\bibfield{author}{\bibinfo{person}{Moataz Abdelaal}, \bibinfo{person}{Nathan~D.
  Schiele}, \bibinfo{person}{Katrin Angerbauer}, \bibinfo{person}{Kuno
  Kurzhals}, \bibinfo{person}{Michael Sedlmair}, {and} \bibinfo{person}{Daniel
  Weiskopf}.} \bibinfo{year}{2023}\natexlab{}.
\newblock \showarticletitle{Comparative Evaluation of Bipartite, Node-Link, and
  Matrix-Based Network Representations}.
\newblock \bibinfo{journal}{\emph{IEEE Transactions on Visualization and
  Computer Graphics}} \bibinfo{volume}{29}, \bibinfo{number}{1}
  (\bibinfo{year}{2023}), \bibinfo{pages}{896--906}.
\newblock
\urldef\tempurl%
\url{https://doi.org/10.1109/TVCG.2022.3209427}
\showDOI{\tempurl}


\bibitem[Beck et~al\mbox{.}(2016)]%
        {beck2016expert}
\bibfield{author}{\bibinfo{person}{Fabian Beck}, \bibinfo{person}{Yasett
  Acurana}, \bibinfo{person}{Tanja Blascheck}, \bibinfo{person}{Rudolf Netzel},
  {and} \bibinfo{person}{Daniel Weiskopf}.} \bibinfo{year}{2016}\natexlab{}.
\newblock \showarticletitle{An expert evaluation of word-sized visualizations
  for analyzing eye movement data}. In \bibinfo{booktitle}{\emph{Workshop on
  Eye Tracking and Visualization}}. \bibinfo{publisher}{Springer},
  \bibinfo{pages}{50--54}.
\newblock
\urldef\tempurl%
\url{https://doi.org/10.1109/ETVIS.2016.7851166}
\showDOI{\tempurl}


\bibitem[Beck et~al\mbox{.}(2015)]%
        {beck_word-sized_2017-1}
\bibfield{author}{\bibinfo{person}{Fabian Beck}, \bibinfo{person}{Tanja
  Blascheck}, \bibinfo{person}{Thomas Ertl}, {and} \bibinfo{person}{Daniel
  Weiskopf}.} \bibinfo{year}{2015}\natexlab{}.
\newblock \showarticletitle{Word-{Sized} {Eye}-{Tracking} {Visualizations}}. In
  \bibinfo{booktitle}{\emph{Workshop on Eye Tracking and Visualization}}.
  \bibinfo{publisher}{Springer}, \bibinfo{pages}{113--128}.
\newblock
\urldef\tempurl%
\url{https://doi.org/10.1007/978-3-319-47024-5_7}
\showDOI{\tempurl}


\bibitem[Ben-Yehudah and Eshet-Alkalai(2018)]%
        {ben2018contrib}
\bibfield{author}{\bibinfo{person}{Gal Ben-Yehudah} {and}
  \bibinfo{person}{Yoram Eshet-Alkalai}.} \bibinfo{year}{2018}\natexlab{}.
\newblock \showarticletitle{The contribution of text-highlighting to
  comprehension: A comparison of print and digital reading}.
\newblock \bibinfo{journal}{\emph{Journal of Educational Multimedia and
  Hypermedia}} \bibinfo{volume}{27}, \bibinfo{number}{2}
  (\bibinfo{year}{2018}), \bibinfo{pages}{153--178}.
\newblock
\urldef\tempurl%
\url{https://www.learntechlib.org/p/174353}
\showURL{%
\tempurl}


\bibitem[Besan{\c{c}}on and Dragicevic(2017)]%
        {besancon_p_confidence_2017}
\bibfield{author}{\bibinfo{person}{Lonni Besan{\c{c}}on} {and}
  \bibinfo{person}{Pierre Dragicevic}.} \bibinfo{year}{2017}\natexlab{}.
\newblock \showarticletitle{The significant difference between p-values and
  confidence intervals}. In \bibinfo{booktitle}{\emph{Conference on
  l’Interaction Homme-Machine}}. \bibinfo{publisher}{Association for
  Computing Machinery}, \bibinfo{pages}{53--62}.
\newblock
\urldef\tempurl%
\url{https://hal.inria.fr/hal-01562281v2}
\showURL{%
\tempurl}


\bibitem[Byrne(1847)]%
        {byrne1847first}
\bibfield{author}{\bibinfo{person}{Oliver Byrne}.}
  \bibinfo{year}{1847}\natexlab{}.
\newblock \bibinfo{booktitle}{\emph{The first six books of the Elements of
  Euclid: in which coloured diagrams and symbols are used instead of letters
  for the greater ease of learners}}.
\newblock \bibinfo{publisher}{William Pickering}.
\newblock


\bibitem[Chat GPT 3.5(2023)]%
        {gpt}
Chat GPT 3.5 \bibinfo{year}{2023}\natexlab{}.
\newblock \bibinfo{title}{{GPT} 3.5 turbo}.
\newblock
\newblock
\newblock
\shownote{platform.openai.com/docs/models/gpt-3-5}.


\bibitem[Chi et~al\mbox{.}(2007)]%
        {chi2007visual}
\bibfield{author}{\bibinfo{person}{Ed~Huai{-}hsin Chi},
  \bibinfo{person}{Michelle Gumbrecht}, {and} \bibinfo{person}{Lichan Hong}.}
  \bibinfo{year}{2007}\natexlab{}.
\newblock \showarticletitle{Visual Foraging of Highlighted Text: An
  Eye-Tracking Study}. In \bibinfo{booktitle}{\emph{12th International
  Conference on {HCI} Intelligent Multimodal Interaction Environments}}.
  \bibinfo{publisher}{Springer}, \bibinfo{pages}{589--598}.
\newblock
\urldef\tempurl%
\url{https://doi.org/10.1007/978-3-540-73110-8\_64}
\showDOI{\tempurl}


\bibitem[Chi et~al\mbox{.}(2005)]%
        {chi2005scent}
\bibfield{author}{\bibinfo{person}{Ed~H. Chi}, \bibinfo{person}{Lichan Hong},
  \bibinfo{person}{Michelle Gumbrecht}, {and} \bibinfo{person}{Stuart~K.
  Card}.} \bibinfo{year}{2005}\natexlab{}.
\newblock \showarticletitle{ScentHighlights: highlighting conceptually-related
  sentences during reading}. In \bibinfo{booktitle}{\emph{10th International
  Conference on Intelligent User Interfaces}}. \bibinfo{publisher}{Association
  for Computing Machinery}, \bibinfo{pages}{272–274}.
\newblock
\urldef\tempurl%
\url{https://doi.org/10.1145/1040830.1040895}
\showDOI{\tempurl}


\bibitem[Cumming(2013)]%
        {cumming2013understanding}
\bibfield{author}{\bibinfo{person}{Geoff Cumming}.}
  \bibinfo{year}{2013}\natexlab{}.
\newblock \bibinfo{booktitle}{\emph{Understanding the new statistics: Effect
  sizes, confidence intervals, and meta-analysis}}.
\newblock \bibinfo{publisher}{Routledge}.
\newblock


\bibitem[Dragicevic(2016)]%
        {dragicevic2016fair}
\bibfield{author}{\bibinfo{person}{Pierre Dragicevic}.}
  \bibinfo{year}{2016}\natexlab{}.
\newblock \bibinfo{booktitle}{\emph{Fair statistical communication in HCI}}.
\newblock \bibinfo{publisher}{Springer}, \bibinfo{pages}{291--330}.
\newblock
\urldef\tempurl%
\url{https://doi.org/10.1007/978-3-319-26633-6_13}
\showDOI{\tempurl}


\bibitem[Fowler and Barker(1974)]%
        {fowler1974effectiveness}
\bibfield{author}{\bibinfo{person}{Robert L~H Fowler} {and}
  \bibinfo{person}{Anne~S. Barker}.} \bibinfo{year}{1974}\natexlab{}.
\newblock \showarticletitle{Effectiveness of highlighting for retention of text
  material}.
\newblock \bibinfo{journal}{\emph{Journal of Applied Psychology}}
  \bibinfo{volume}{59}, \bibinfo{number}{3} (\bibinfo{year}{1974}),
  \bibinfo{pages}{358--364}.
\newblock
\urldef\tempurl%
\url{https://doi.org/10.1037/h0036750}
\showDOI{\tempurl}


\bibitem[Goffin et~al\mbox{.}(2017)]%
        {goffin_exploratory_2017}
\bibfield{author}{\bibinfo{person}{Pascal Goffin}, \bibinfo{person}{Jeremy
  Boy}, \bibinfo{person}{Wesley Willett}, {and} \bibinfo{person}{Petra
  Isenberg}.} \bibinfo{year}{2017}\natexlab{}.
\newblock \showarticletitle{An {Exploratory} {Study} of {Word}-{Scale}
  {Graphics} in {Data}-{Rich} {Text} {Documents}}.
\newblock \bibinfo{journal}{\emph{IEEE transactions on visualization and
  computer graphics}} \bibinfo{volume}{23}, \bibinfo{number}{10}
  (\bibinfo{year}{2017}), \bibinfo{pages}{2275--2287}.
\newblock
\urldef\tempurl%
\url{https://doi.org/10.1109/TVCG.2016.2618797}
\showDOI{\tempurl}


\bibitem[Goffin et~al\mbox{.}(2015)]%
        {goffin_exploring_2015}
\bibfield{author}{\bibinfo{person}{Pascal Goffin}, \bibinfo{person}{Wesley
  Willett}, \bibinfo{person}{Anastasia Bezerianos}, {and}
  \bibinfo{person}{Petra Isenberg}.} \bibinfo{year}{2015}\natexlab{}.
\newblock \showarticletitle{Exploring the effect of word-scale visualizations
  on reading behavior}. In \bibinfo{booktitle}{\emph{{Conference} on {Human}
  {Factors} in {Computing} {Systems}}}. \bibinfo{publisher}{Association for
  Computing Machinery}, \bibinfo{pages}{1827--1832}.
\newblock
\urldef\tempurl%
\url{https://doi.org/10.1145/2702613.2732778}
\showDOI{\tempurl}


\bibitem[Goldberg and Helfman(2011)]%
        {goldberg2011}
\bibfield{author}{\bibinfo{person}{Joseph Goldberg} {and}
  \bibinfo{person}{Jonathan Helfman}.} \bibinfo{year}{2011}\natexlab{}.
\newblock \showarticletitle{Eye tracking for visualization evaluation: Reading
  values on linear versus radial graphs}.
\newblock \bibinfo{journal}{\emph{Information Visualization}}
  \bibinfo{volume}{10}, \bibinfo{number}{3} (\bibinfo{year}{2011}),
  \bibinfo{pages}{182--195}.
\newblock
\urldef\tempurl%
\url{https://doi.org/10.1177/1473871611406623}
\showDOI{\tempurl}


\bibitem[Huey(1908)]%
        {huey1908}
\bibfield{author}{\bibinfo{person}{Edmund~Burke Huey}.}
  \bibinfo{year}{1908}\natexlab{}.
\newblock \bibinfo{booktitle}{\emph{The psychology and pedagogy of reading:
  With a review of the history of reading and writing and of methods, texts,
  and hygiene in reading}}.
\newblock \bibinfo{publisher}{Macmillan}.
\newblock


\bibitem[Huth et~al\mbox{.}(2021a)]%
        {huth2021online}
\bibfield{author}{\bibinfo{person}{Franziska Huth}, \bibinfo{person}{Miriam
  Awad-Mohammed}, \bibinfo{person}{Johannes Knittel}, \bibinfo{person}{Tanja
  Blascheck}, {and} \bibinfo{person}{Petra Isenberg}.}
  \bibinfo{year}{2021}\natexlab{a}.
\newblock \showarticletitle{Online Study of Word-Sized Visualizations in Social
  Media}. In \bibinfo{booktitle}{\emph{EuroVis 2021 - Posters}}.
  \bibinfo{publisher}{The Eurographics Association}, \bibinfo{pages}{13--15}.
\newblock
\urldef\tempurl%
\url{https://doi.org/10.2312/evp.20211069}
\showDOI{\tempurl}


\bibitem[Huth et~al\mbox{.}(2021b)]%
        {huth_word-sized_2021}
\bibfield{author}{\bibinfo{person}{Franziska Huth}, \bibinfo{person}{Tanja
  Blascheck}, \bibinfo{person}{Steffen Koch}, \bibinfo{person}{Sonja Utz},
  {and} \bibinfo{person}{Thomas Ertl}.} \bibinfo{year}{2021}\natexlab{b}.
\newblock \showarticletitle{Word-sized {Visualizations} for {Exploring}
  {Discussion} {Diversity} in {Social} {Media}}. In
  \bibinfo{booktitle}{\emph{IVAPP 2021-12th International Conference on
  Computer Vision, Imaging and Computer Graphics Theory and Applications}}.
  \bibinfo{publisher}{SCITEPRESS}, \bibinfo{pages}{256--265}.
\newblock
\urldef\tempurl%
\url{https://doi.org/10.5220/0010328602560265}
\showDOI{\tempurl}


\bibitem[Krzywinski and Altman(2013)]%
        {krzywinski2013error}
\bibfield{author}{\bibinfo{person}{Martin Krzywinski} {and}
  \bibinfo{person}{Naomi Altman}.} \bibinfo{year}{2013}\natexlab{}.
\newblock \showarticletitle{Error bars: the meaning of error bars is often
  misinterpreted, as is the statistical significance of their overlap}.
\newblock \bibinfo{journal}{\emph{Nature methods}} \bibinfo{volume}{10},
  \bibinfo{number}{10} (\bibinfo{year}{2013}), \bibinfo{pages}{921--923}.
\newblock
\urldef\tempurl%
\url{https://doi.org/10.1038/nmeth.2659}
\showDOI{\tempurl}


\bibitem[Latif and Beck(2018)]%
        {latif_visually_2018}
\bibfield{author}{\bibinfo{person}{Shahid Latif} {and} \bibinfo{person}{Fabian
  Beck}.} \bibinfo{year}{2018}\natexlab{}.
\newblock \showarticletitle{Visually {Augmenting} {Documents} {With} {Data}}.
\newblock \bibinfo{journal}{\emph{Computing in Science Engineering}}
  \bibinfo{volume}{20}, \bibinfo{number}{6} (\bibinfo{year}{2018}),
  \bibinfo{pages}{96--103}.
\newblock
\urldef\tempurl%
\url{https://doi.org/10.1109/MCSE.2018.2875316}
\showDOI{\tempurl}


\bibitem[Lorch(1989)]%
        {lorch1989text}
\bibfield{author}{\bibinfo{person}{Robert~F. Lorch}.}
  \bibinfo{year}{1989}\natexlab{}.
\newblock \showarticletitle{Text-signaling devices and their effects on reading
  and memory processes}.
\newblock \bibinfo{journal}{\emph{Educational Psychology Review}}
  \bibinfo{volume}{1} (\bibinfo{year}{1989}), \bibinfo{pages}{209--234}.
\newblock
\urldef\tempurl%
\url{https://doi.org/10.1007/BF01320135}
\showDOI{\tempurl}


\bibitem[Netzel et~al\mbox{.}(2017)]%
        {netzel2017}
\bibfield{author}{\bibinfo{person}{Rudolf Netzel}, \bibinfo{person}{Bettina
  Ohlhausen}, \bibinfo{person}{Kuno Kurzhals}, \bibinfo{person}{Robin Woods},
  \bibinfo{person}{Michael Burch}, {and} \bibinfo{person}{Daniel Weiskopf}.}
  \bibinfo{year}{2017}\natexlab{}.
\newblock \showarticletitle{User performance and reading strategies for metro
  maps: An eye tracking study}.
\newblock \bibinfo{journal}{\emph{Spatial Cognition \& Computation}}
  \bibinfo{volume}{17}, \bibinfo{number}{1-2} (\bibinfo{year}{2017}),
  \bibinfo{pages}{39--64}.
\newblock
\urldef\tempurl%
\url{https://doi.org/10.1080/13875868.2016.1226839}
\showDOI{\tempurl}


\bibitem[OSF(2024)]%
        {osf}
OSF \bibinfo{year}{2024}\natexlab{}.
\newblock \bibinfo{title}{{OSF}--{E}ye {T}racking on {T}ext {R}eading with
  {V}isual {E}nhancements}.
\newblock
\newblock
\newblock
\shownote{\url{https://osf.io/utw7b/}}.


\bibitem[Rayner(1977)]%
        {rayner1977}
\bibfield{author}{\bibinfo{person}{Keith Rayner}.}
  \bibinfo{year}{1977}\natexlab{}.
\newblock \showarticletitle{Visual attention in reading: Eye movements reflect
  cognitive processes}.
\newblock \bibinfo{journal}{\emph{Memory \& Cognition}} \bibinfo{volume}{5},
  \bibinfo{number}{4} (\bibinfo{year}{1977}), \bibinfo{pages}{443--448}.
\newblock
\urldef\tempurl%
\url{https://doi.org/10.3758/BF03197383}
\showDOI{\tempurl}


\bibitem[Rayner(1998)]%
        {rayner1998eye}
\bibfield{author}{\bibinfo{person}{Keith Rayner}.}
  \bibinfo{year}{1998}\natexlab{}.
\newblock \showarticletitle{Eye movements in reading and information
  processing: 20 years of research}.
\newblock \bibinfo{journal}{\emph{Psychological Bulletin}}
  \bibinfo{volume}{124}, \bibinfo{number}{3} (\bibinfo{year}{1998}),
  \bibinfo{pages}{372--422}.
\newblock
\urldef\tempurl%
\url{https://doi.org/10.1037/0033-2909.124.3.372}
\showDOI{\tempurl}


\bibitem[Ruchikachorn and Rattanawicha(2018)]%
        {ruchikachorn2018eye}
\bibfield{author}{\bibinfo{person}{Puripant Ruchikachorn} {and}
  \bibinfo{person}{Pimmanee Rattanawicha}.} \bibinfo{year}{2018}\natexlab{}.
\newblock \showarticletitle{An Eye-Tracking Study on Sparklines within Textual
  Context.}. In \bibinfo{booktitle}{\emph{EuroVis 2018 - Posters}}.
  \bibinfo{publisher}{The Eurographics Association}, \bibinfo{pages}{17--19}.
\newblock
\urldef\tempurl%
\url{https://doi.org/10.2312/eurp.20181119}
\showDOI{\tempurl}


\bibitem[Tufte(2006)]%
        {tufte_beautiful_2006}
\bibfield{author}{\bibinfo{person}{Edward~R. Tufte}.}
  \bibinfo{year}{2006}\natexlab{}.
\newblock \bibinfo{booktitle}{\emph{Beautiful {Evidence}} (\bibinfo{edition}{1}
  ed.)}.
\newblock \bibinfo{publisher}{Graphics Press}.
\newblock


\bibitem[Von~Restorff(1933)]%
        {restorff1933wirkung}
\bibfield{author}{\bibinfo{person}{Hedwig Von~Restorff}.}
  \bibinfo{year}{1933}\natexlab{}.
\newblock \showarticletitle{{\"U}ber die Wirkung von Bereichsbildungen im
  Spurenfeld}.
\newblock \bibinfo{journal}{\emph{Psychologische Forschung}}
  \bibinfo{volume}{18} (\bibinfo{year}{1933}), \bibinfo{pages}{299--342}.
\newblock
\urldef\tempurl%
\url{https://doi.org/10.1007/BF02409636}
\showDOI{\tempurl}


\bibitem[Yeari et~al\mbox{.}(2017)]%
        {yeari2017effect}
\bibfield{author}{\bibinfo{person}{Menahem Yeari}, \bibinfo{person}{Marja
  Oudega}, {and} \bibinfo{person}{Paul van~den Broek}.}
  \bibinfo{year}{2017}\natexlab{}.
\newblock \showarticletitle{The effect of highlighting on processing and memory
  of central and peripheral text information: evidence from eye movements}.
\newblock \bibinfo{journal}{\emph{Journal of Research in Reading}}
  \bibinfo{volume}{40}, \bibinfo{number}{4} (\bibinfo{year}{2017}),
  \bibinfo{pages}{365--383}.
\newblock
\urldef\tempurl%
\url{https://doi.org/10.1111/1467-9817.12072}
\showDOI{\tempurl}


\end{thebibliography}

\end{document}